\documentclass[review]{elsarticle}
\usepackage[utf8]{inputenc}
\usepackage{subfig}
\usepackage{multirow}
\usepackage{amsmath}
\usepackage{longtable}
\usepackage{changepage}
\usepackage{algorithm,algorithmic}
\usepackage[T1]{fontenc}
\usepackage{rotating}
\usepackage{hyperref}
\hypersetup{
    colorlinks=true,
    linkcolor=red,
    filecolor=magenta,      
    urlcolor=blue,
    citecolor=green,
}
\usepackage{soul}
\usepackage[table,xcdraw]{xcolor}
\newcommand{\orcid}[1]{\href{https://orcid.org/#1}{\includegraphics[width=10pt]{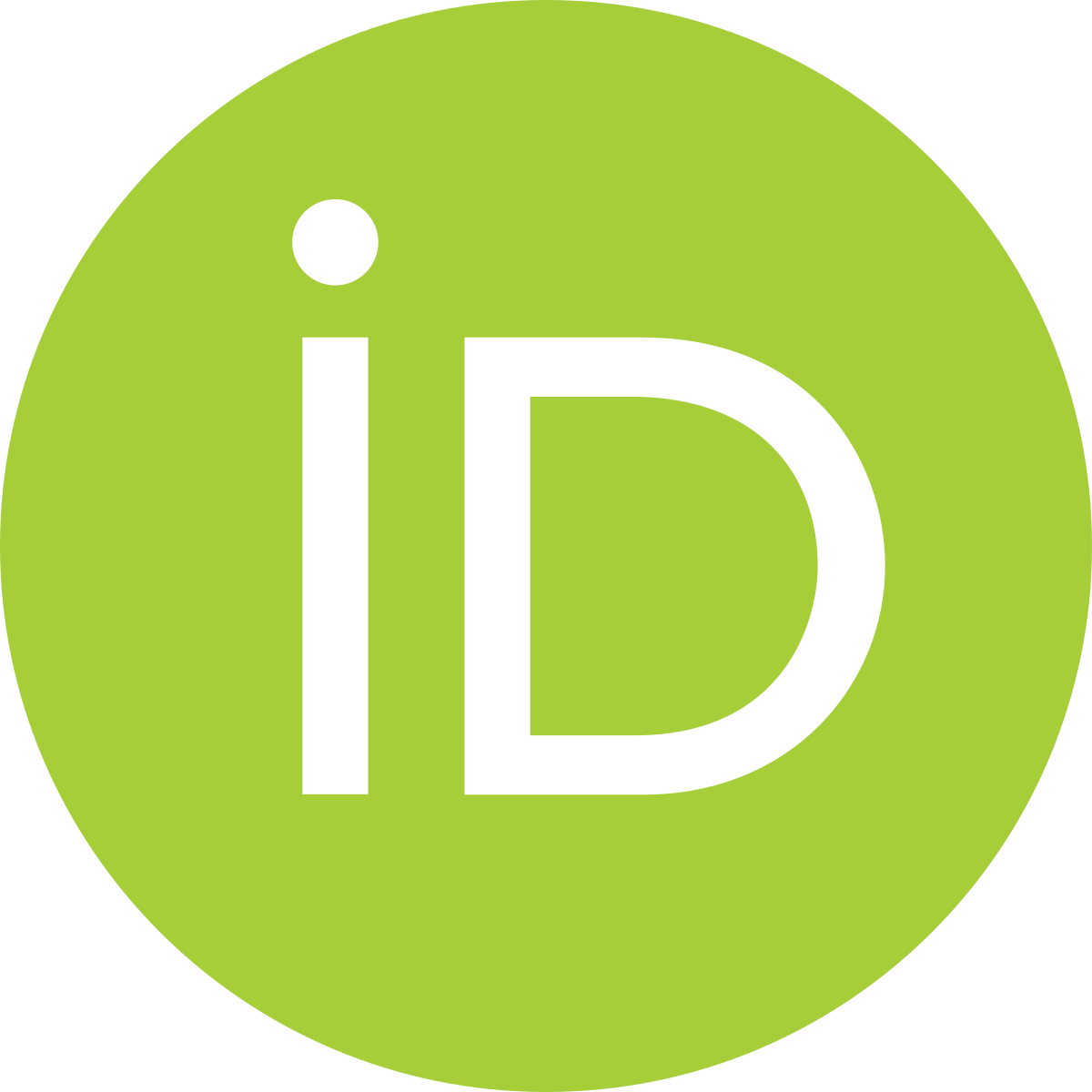}}}

\begin{document}

\begin{frontmatter}

\title{MFSNet: A Multi Focus Segmentation Network for Skin Lesion Segmentation}

\author[a]{Hritam Basak \orcid{0000-0001-5921-1230}}
\author[a]{Rohit Kundu\orcid{0000-0001-8665-8898}\corref{cor1}}
\cortext[cor1]{Corresponding author}
\ead{rohitkunduju@gmail.com}
\author[b]{Ram Sarkar\orcid{0000-0001-8813-4086}}

\address[a]{Department of Electrical Engineering, Jadavpur University, INDIA}
\address[b]{Department of Computer Science \& Engineering, Jadavpur University, INDIA}

\begin{abstract} 
Segmentation is essential for medical image analysis to identify and localize diseases, monitor morphological changes, and extract discriminative features for further diagnosis. Skin cancer is one of the most common types of cancer globally, and its early diagnosis is pivotal for the complete elimination of malignant tumors from the body. This research develops an Artificial Intelligence (AI) framework for supervised skin lesion segmentation employing the deep learning approach. The proposed framework, called MFSNet (Multi-Focus Segmentation Network), uses differently scaled feature maps for computing the final segmentation mask using raw input RGB images of skin lesions. In doing so, initially, the images are preprocessed to remove unwanted artifacts and noises. The MFSNet employs the Res2Net backbone, a recently proposed convolutional neural network (CNN), for obtaining deep features used in a Parallel Partial Decoder (PPD) module to get a global map of the segmentation mask. In different stages of the network, convolution features and multi-scale maps are used in two boundary attention (BA) modules and two reverse attention (RA) modules to generate the final segmentation output. MFSNet, when evaluated on three publicly available datasets: $PH^2$, ISIC 2017, and HAM10000, outperforms state-of-the-art methods, justifying the reliability of the framework. The relevant codes for the proposed approach are accessible at \url{https://github.com/Rohit-Kundu/MFSNet}
\end{abstract}

\end{frontmatter}

\section{Introduction}\label{intro}

\begin{figure}
    \centering
    \subfloat[Raw dermoscopy image]{\includegraphics[scale=0.14]{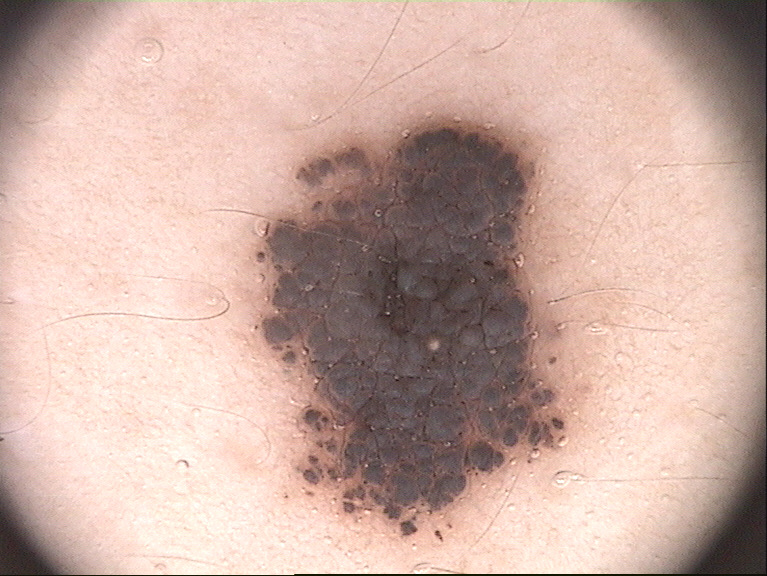}}\;\;
    \subfloat[Ground truth mask]{\includegraphics[scale=0.14]{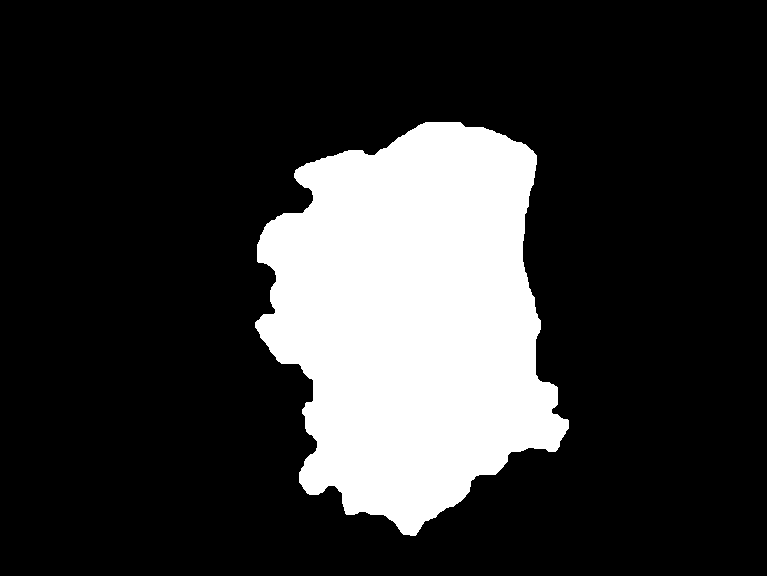}}
    \caption{Example of a skin lesion image and its ground truth mask.}
    \label{example}
\end{figure}

Melanoma is the most severe and deadly type of skin cancer, causing more than 13 thousand incidences globally. Though less prevalent than its non-malignant counterpart, malignant melanoma is increasing at an alarming rate of 4\% per year. Research has shown its correlation with genetic and physical variations. The primary cause of melanoma is long-term exposure to ultraviolet (UV) rays. With the increase in greenhouse gases, the protective ozone layer in the stratosphere is depleting rapidly, causing the harmful solar UV rays to reach the earth's surface. This causes the global incidence of melanoma to rise rapidly. Fortunately, studies like Siegel et al. \cite{siegel2019cancer} show that early detection can decrease the chances of fatality by 97\%. Surgical treatment of melanoma is often disfiguring and extremely painful, justifying the importance of early detection of the disease. Dermoscopy is a non-invasive test for detecting and diagnosing pigmented skin lesions and malignant melanoma in the early stages. It is often considered the golden standard for melanoma localization. However, manual labeling and reviewing are extremely grueling and cumbersome even for expert clinicians, relying on their perceptions and vision. Therefore, to mitigate the problem, Computer-Aided Diagnosis (CAD) systems have been widely preferred as a support system to aid clinicians in automated segmentation and analysis of malignant melanoma.

\textit{Semantic segmentation} refers to the pixel-level classification of the images. Each pixel in an image is classified as part of the object class or background class. This is beneficial for localizing the region of interest (ROI) from the raw images for further analysis and thus is a vital preprocessing step in automated disease diagnosis. \autoref{example}(a) shows an example of a raw skin-lesion image. Its segmented image, called ``ground truth," is shown in \autoref{example}(b). Here, the image is classified into two classes, namely ``lesion" and ``background," which led to the generation of a ``binary mask" image. The task of semantic segmentation is to generate a segmentation map like \autoref{example}(b) from raw input image similar to \autoref{example}(a). To address this, extensive research attempts have been made since the last decade to automate the segmentation of lesions, monitor their growth, and aid physicians in making surgical decisions, thereby increasing the clinical significance.

Segmentation of skin melanoma from non-invasive dermoscopy images relies upon several emerging and traditional methods. Among them, segmentation methods based on artificial intelligence have widely been explored and adopted due to their excellent accuracy, robustness, and reliability. Extensive research has been conducted in the last few years, using neural networks \cite{attia2017skin}, fuzzy logic \cite{garcia2019segmentation}, attention-gated networks \cite{wang2019automated}, or their combinations with traditional image processing methods to improve the segmentation performance. The significant variations in texture, size, shape, the position of lesions, and obscure boundaries in dermoscopy images make it extremely challenging to obtain accurate and prominent tissue-level segmentation maps for developing CAD systems. Pre and post-processing are the other essential aspects and used in most of the current segmentation methods \cite{chatterjee2019integration} for effectively removing artifacts, enhancing image quality, removing unnecessary noises in images for effective and accurate segmentation of pigmented skin lesions from images. Beuren et al. \cite{beuren2012skin} proposed a series of morphological operations for image enhancement of image resolution and denoising before segmentation. Later Chatterjee et al. \cite{chatterjee2019integration} proposed the Fractal Region Texture Analysis (FRTA) method for quantification of texture information integrated with Recursive Feature Elimination (RFE) and several morphological operations as preprocessing before classification of dermoscopic images. Verma et al. \cite{verma2015enhancement} showed that median filters and anisotropic diffusion filters can be helpful in not only smoothing the images but also removal of thick hairlines, preserving sufficient lesion edge information. Recently, morphological operations and image inpainting methods have been modified and used in research for dermoscopy image analysis \cite{salido2018using}. In the preprocessing step, this research has incorporated the image inpainting method for unwanted hair removal from the input images. 

In literature, the skin lesion segmentation methods are broadly classified into the following categories: (a) edge detection and thresholding \cite{ma2015novel}, (b) active contour models \cite{vasconcelos2019automatic}, and (c) segmentation based on convolutional neural network (CNN) \cite{wei2019attention, basak2020comparative}, etc. Symmetrical encoder-decoder architecture, also known as U-Net, proposed by Ronneberger et al. \cite{ronneberger2015u}, is widely used and considered as the golden standard for several image segmentation tasks. It consists of a downsampling path that captures sufficient semantics and context, connected to an expanding path for accurate localization of the ROI. Later Zhou et al. \cite{zhou2018unet++} proposed a novel architecture UNet++ by redesigning the series of nested dense skip connections to reduce the semantic gap between the feature representations and the encoder-decoder sub-networks. Their proposed model outperformed the previous U-Net architecture in multiple biomedical image segmentation tasks. Weng et al. \cite{weng2019unet} proposed another modification in the U-Net backbone by incorporating neural architecture search (NAS), thereby improving the segmentation performance significantly. SegNet \cite{badrinarayanan2017segnet} is a similar encoder-decoder model, for instance, segmentation, that uses a VGG16 backbone followed by a decoder path integrated with a pixel-wise classification layer. This is a well-known segmentation model for binary or multi-class segmentation problems and has been proven to produce state-of-the-art results in various domains. Yuan et al. \cite{yuan2017automatic} proposed a fully convolution-deconvolution network that was able to produce a dice similarity score of 76.5\% on the ISIC 2017 dataset. Later Abraham et al. \cite{abraham2019novel} proposed a novel Focal Tversky Loss function, then integrated with attention U-Net, produced a state-of-the-art result on BUS 2017B and ISIC2018 datasets with average dice scores of 80.4\% and 85.6\% respectively. Double U-Net \cite{jha2020doubleu}, another modification of U-Net that used two different upsampling branches instead of one, was used to produce two different segmentation maps, slightly different from one another. The paper reported an average dice score of 89.2\% on the ISIC 2017 dataset.

Recently meta-heuristic-based optimization algorithms have been explored for thresholding-based segmentation and image enhancement operations in different applications. Aljanabi et al. \cite{aljanabi2018skin} proposed an image thresholding method by selecting an optimum threshold level using the artificial bee colony (ABC) algorithm. The algorithm was able to produce segmentation maps with high confidence on several widely known skin datasets. Attention mechanisms are also widely known for boosting the performance of CNN-based models in different computer vision applications. Chattopadhyay et al. \cite{chattopadhyay2020multi} proposed a multi-scale attention mechanism which is inspired by the work of \cite{bi2020multi} for accurate localization and segmentation of objects. The dual attention mechanism was proposed by \cite{fu2019dual} adaptively integrates the local features with their corresponding global dependencies. Though used in scene segmentation application, it inspired several similar works in the biomedical domain \cite{barata2021explainable}. Generative Adversarial Networks (GAN) have also been instrumental for extensive research for biomedical image segmentation recently \cite{lei2020skin}.

\subsection{Overview and Contributions}
To address the issues mentioned before, we propose a novel skin lesion segmentation framework, called Multi-Focus Segmentation Network (MFSNet), that produces the final segmentation map by focusing on image information at multiple scales. Taking a clue from the standard clinical practice, we can say that the area and boundary are the two essential aspects to produce the accurate pixel-level segmentation map based on local appearance from a coarse localization of the melanoma region. The proposed model generates a coarse segmentation map implicitly by aggregating image features at multiple levels, followed by a series of reverse and boundary attention networks by iteratively learning pixel-level information of area and boundary by explicitly using the coarse map and ground truth the global guidance. We have evaluated the performance of the proposed model on three publicly available skin melanoma datasets: The $PH^2$ dataset by Mendoncca et al. \cite{mendoncca2013ph}, the ISIC 2017 dataset by Codella et al. \cite{codella2018skin} and the HAM10000 dataset by Tschandl et al. \cite{tschandl2018ham10000}. The proposed model outperforms state-of-the-art models on the same datasets justifying the reliability and robustness of the framework.

The contributions of the present research are as follows:
\begin{enumerate}
    \item The use of differently focused segmentation maps in various stages of the proposed MFSNet helps accurately map both the lesion's coarse structure and its fine edges.
    \item \textcolor{black}{Unlike the commonly used segmentation frameworks in literature, the proposed model upsamples the encoded features in subsequent steps of attention modules instead of coarse upsampling applied in U-Net type architectures.}
    \item We evaluate the proposed MFSNet model on three publicly available datasets: $PH^2$, ISIC 2017 and HAM10000 datasets, and obtain dice similarity coefficient values of 0.954, 0.987, and 0.906 respectively on the datasets, outperforming state-of-the-art methods on the same datasets.
\end{enumerate}

\section{Proposed Method}\label{method}
This section describes the architecture of our proposed MFSNet, which combines the high-level semantics and the low-level edge information by using a series of RA modules, BA block, and a PPD module. We propose a hybrid loss function that integrates the widely used Binary Cross-Entropy (BCE) loss with the Weighted IoU loss functions. The whole segmentation process is followed by image inpainting and a preprocessing step for artifact removal, described in Section \ref{preprocess}.

\subsection{Image Preprocessing}\label{preprocess}

\begin{figure*}
    \centering
    \resizebox{\textwidth}{!}{
    \subfloat[Original Image]{\includegraphics[scale=0.5]{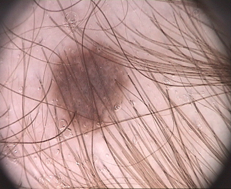}}\;\;
    \subfloat[Grayscale Image]{\includegraphics[scale=0.5]{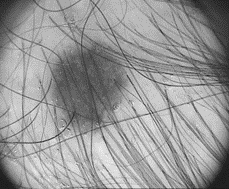}}\;\;
    \subfloat[Blackhat Filtered Image]{\includegraphics[scale=0.5]{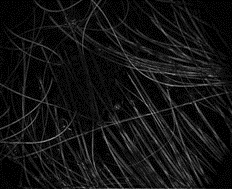}}\;\;
    \subfloat[Threshold for Inpainting]{\includegraphics[scale=0.5]{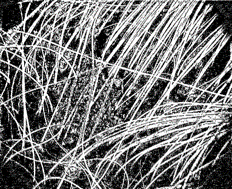}}\;\;
    \subfloat[Final Inpainted Image]{\includegraphics[scale=0.5]{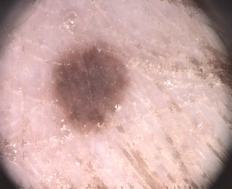}}
    }\\
    \resizebox{\textwidth}{!}{
    \subfloat[Original Image]{\includegraphics[scale=0.5]{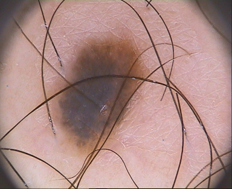}}\;\;
    \subfloat[Grayscale Image]{\includegraphics[scale=0.5]{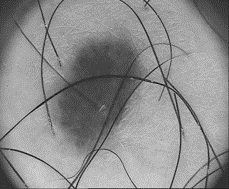}}\;\;
    \subfloat[Blackhat Filtered Image]{\includegraphics[scale=0.5]{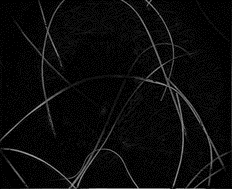}}\;\;
    \subfloat[Threshold for Inpainting]{\includegraphics[scale=0.5]{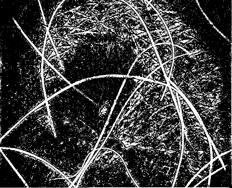}}\;\;
    \subfloat[Final Inpainted Image]{\includegraphics[scale=0.5]{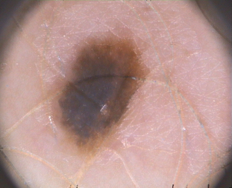}}
    }
    \caption{Outputs of the image inpainting method used for artefact removal on the PH2 dataset: (a) \& (f)- Original images; (b) \& (g)- Corresponding grayscale images; (c) \& (h)- Blackhat filtered images; (d) \& (i)- Thresholding for the inpainting operation; (e) \& (j)- final preprocessed (inpainted) images.}
    \label{inpaint}
\end{figure*}


Dermoscopy images vary in terms of size, pixel intensity and may suffer from unwanted artifacts in the form of noises or body hair. These artifacts may lead to abrupt segmentation results in some images and may diminish the overall model performance. Hence, to address these problems, we used standard image preprocessing methods before segmenting the images. All the images have been resized to a shape of $256\times 256$ for faster convolution and resolving excessive memory constraints. Next, we perform image normalization to resolve the uneven image contrast issues. Finally, we introduce the image inpainting method for hair removal.

Following the work of \cite{telea2004image}, we have used several morphological operations for hair removal from dermoscopy images. First, the input RGB images are converted to grayscale images, followed by blackhat transformation as proposed by \cite{wang2014morphological}. In this regard, we define a structuring element: a cross-shaped two-dimensional array of shape $17\times 17$, i.e., an array whose middle row and column are composed of $1$'s and all other places contain $0$. 



Similar to \cite{wang2014morphological}, closing is also performed to remove small hollows inside a region while keeping the original region shape and size unaltered. Thus the blackhat transformation results in an output image containing elements darker than the surrounding pixel values, whereas smaller than the structuring element. A suitable threshold value is applied to the obtained output from the blackhat transformation to obtain the hair-like artifacts. 

Fast marching method \cite{long2015fully} is widely used for segmentation purposes. In this research, we have used this algorithm for image inpainting. We used the thresholded image output from blackhat transformation and the original input image and replaced the artifacts or hair structures with the neighboring pixels. \autoref{inpaint} shows the image outputs from different intermediate steps of image artifact removal.

\subsection{MFSNet Architecture}\label{architecture}
\begin{figure*}[tbp]
    \centering
    \resizebox{\textwidth}{!}{
    \includegraphics[scale=0.5]{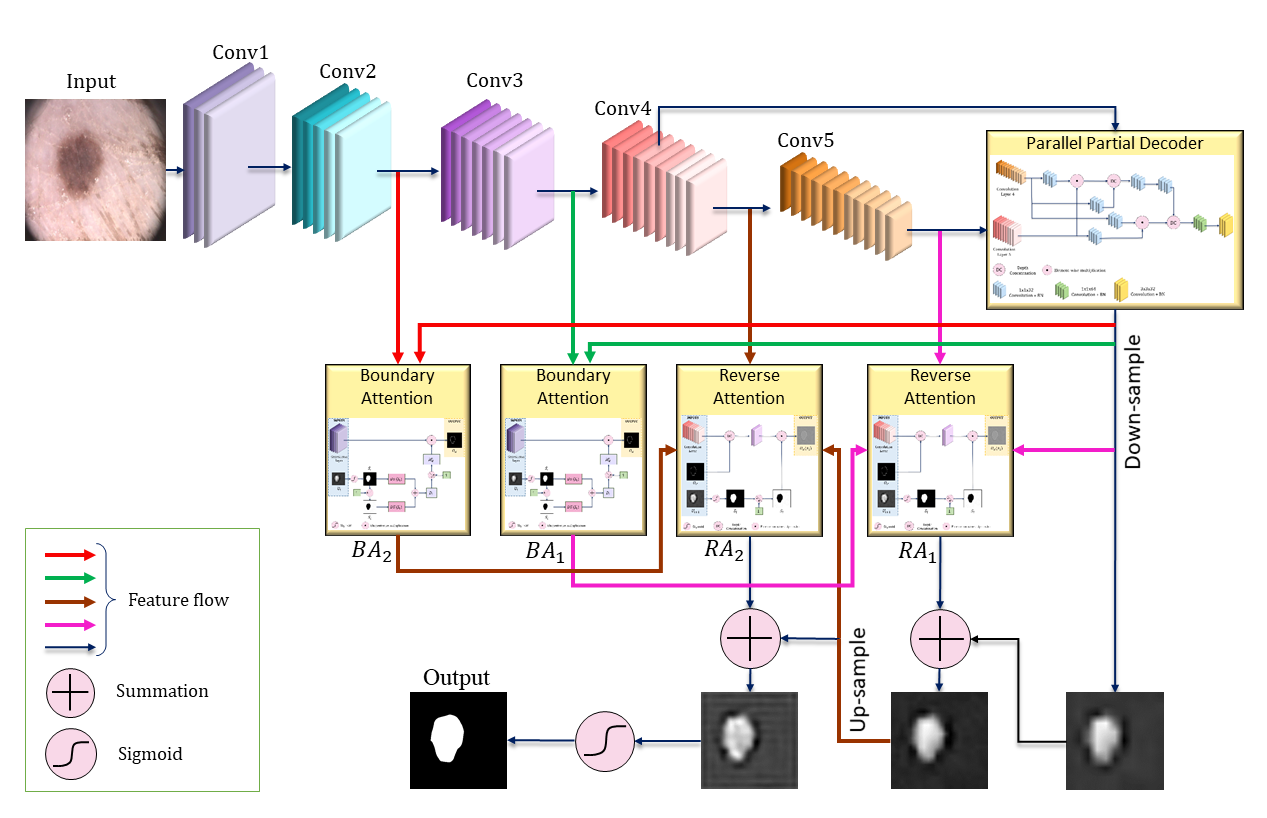}
    }
    \caption{\textcolor{black}{Overall structure of the proposed MFSNet model for the segmentation of skin lesions. The inputs to the boundary attention and reverse attention blocks have been shown in different colored arrows.} The inputs of $BA_1$, $BA_2$, $RA_1$, and $RA_2$ are marked using green, red, pink and brown arrows respectively.}
    \label{overall}
\end{figure*}

\autoref{overall} shows the architecture of the proposed MFSNet. It consists of the Res2Net as a backbone, which is a recently proposed CNN model \cite{gao2019res2net}, for feature extraction combined with a series of RA branches, explained in Section \ref{RA}. Only five initial convolution layers of the network are used for this purpose. The first three layers are used to extract low-level features with high resolution but very little spatial information. The second and third level features, $F_2$ and $F_3$, with important edge information, are fed to the BA module to improve melanoma boundary representation. $F_2$ and $F_3$ are further used for two different purposes. They are fed to the following two layers of the CNN, whose output is fed to the PPD module to generate the global segmentation map $O_S$, which is used as the global map for coarse localization of melanoma segmentation. Secondly, they are fed to the RA branches, along with $O_S$, to be used as the global guidance for the entire learning process of the network. The subsequent two layers of features $F_i; i={4, 5}$ from the successive two consecutive layers of the CNN are fed to the corresponding RA module to produce output $O_R(F_i)$,\textcolor{black}{ which is concatenated with the upsampled $O_R(F_{i+1})$ from the next branch}, thereby ensuring the multi-level feature representation. This results in the output $O_i$ from each branch, which is supervised with the ground truth $G$ through a loss function (described in Section \ref{loss}). Thus, by using the parallel RA and the residual connections between the segmentation of multiple scales and the ground truth, the errors can be removed by "larger-scale adaptability" \cite{chen2018reverse}. Finally, the output $O_2$ is passed through the sigmoid activation function to produce the final segmentation map $S$. 

In general, the error between the input and output of the RA unit is minor (zero in the extreme case), thus making the learning comparatively easy with very few parameters. Hence, the network can be very effective in region segmentation with fewer parameters. As the learning procedure of the network is focused on generating multiple levels of outputs from multiple branches, the network is named as MFSNet.

\subsection{Workflow of the MFSNet}\label{featflow}
As shown in \autoref{overall}, the proposed model consists of a series of convolution operations, RA branches, and BA modules. For the convenience of the readers, we have described below the flow of information from the input image through the layers and branches to produce the segmentation output finally.
\begin{enumerate}
    \item \textcolor{black}{The input image is initially passed through a series of convolution layers for feature extraction using the Res2Net backbone, where downsampling is performed}. Among those, only the features of the second and third Convolution layers are considered useful for edge guidance of the learning process because the low-level features preserve sufficient boundary information \cite{zhang2019net}. Hence they are used for the BA module that explicitly learns the boundary information. \textcolor{black}{Upsampling is performed in the PPD module. }
    \item The BA module simultaneously takes input from the global segmentation map (output from PPD) and the shallow features from the convolution layers. By performing a series of distance transformations and other mathematical operations, an enhanced boundary map is obtained, further used by the RA modules. The detailed algorithm and workflow of BA are described later in Section \ref{boundary}.
    \item The RA module takes the features from the corresponding convolution layer, BA module, and the upsampled segmentation map from the next layer. The RA module uses two separate input branches to learn features to produce segmentation masks associated with two different classes - foreground and background. Thus the RA module generates a per-class mask to amplify the reverse-class response in the regions that contain high-level semantic information shared between two adjacent classes. Finally, the prediction of these two branches is fused to generate the segmentation output from the RA branch. The detailed workflow of the RA module is explained in Section \ref{RA}.
\end{enumerate}

\subsection{Parallel Reverse Attention branch}\label{RA}

\begin{figure*}
    \centering
    \includegraphics[scale=0.5]{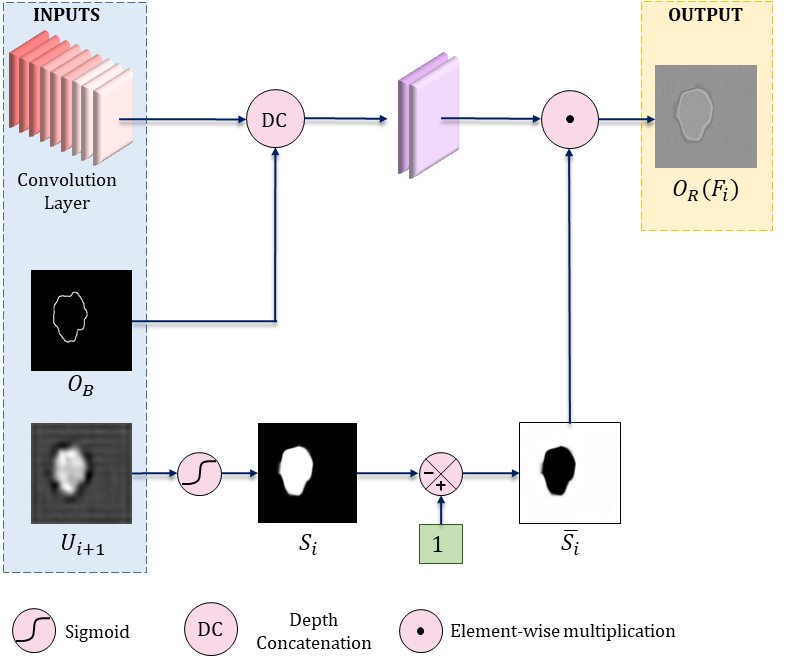}
    \caption{\textcolor{black}{Architecture of the RA module used in the proposed MFSNet model}.\\$O_B$: Output from the BA module; $U_{i+1}$: Upsampled output from the next layer; $O_R(F_i)$: Output of the RA block.}
    \label{ra_fig}
\end{figure*}

In medical diagnosis, clinicians go for a rough estimation of skin melanoma before looking into the tissue-level finer details for proper localization and labeling. Though, it is not easy for a network to learn residual refinement for saliency detection without proper supervision, leading to inaccurate segmentation results. As most of the existing methods heavily rely on image classification networks, fine-tuned for responsiveness to very few discriminatory regions in images, it deviates from the requirement of exploration of pixel-wise prediction of dense regions. We propose a two-stage segmentation method using a parallel RA unit to mitigate this problem and replicate the real-world clinical approach. The deep layers of the CNN produce coarse-level and a rough estimation of the melanoma region, with small structural details \cite{burdick2018rethinking}. Next, followed by the idea of progressive erasing of the foreground region \cite{wei2017object}, we mine discriminative melanoma regions using the RA unit. Instead of aggregating features from all the CNN layers, \cite{chen2018reverse}, our proposed RA model guides the learning of the whole network, starting from the coarse saliency map produced by the deepest CNN layer, containing the highest semantic confidence, by sequentially discovering new information about complementary melanoma regions from the side-output of the last three layers only. \autoref{ra_fig} shows the architecture of the RA module used in the proposed MFSNet.

Let us consider the last two layers of the CNN have features  $F_i; i=\{4,5\}$, the BA module has output $O_B$, and the RA mask of the $i^{th}$ level is $\mathcal{M}_{RA}^i$, then the RA output $O_{R}$ of the $i^{th}$ level is given by \autoref{eqra},\textcolor{black}{ where $\mathcal{D}$ is the downsampling operation,} $\oplus$ is the concatenation operation of downsampled $O_B$ with the $i^{th}$ level feature set $F_i$, $Con$ is the operation of passing the feature through a couple of convolutional layers with filter size set to 64, and $\odot$ is the element-wise multiplication of the concatenated feature with the RA mask $\mathcal{M}_{RA}$.

\begin{equation}\label{eqra}
    O_{R}(F_i)=\mathcal{M}_{RA}^i\odot Con[F_i\oplus \mathcal{D}(O_B)],
\end{equation}

Chen et al. \cite{chen2018reverse} defined the RA mask as in \autoref{ra_mask}, where $\epsilon$ is the operation of forming a 64-channel tensor by repeating the single-channel output, to match the dimension, $\ominus$ is the subtraction operation, $Softmax$ indicates the sigmoid activation, $S_{i+1}$ is the segmentation mask obtained from the $(i+1)^{th}$ layer of the CNN, \textcolor{black}{$\mathcal{U}$ is the upsampling operation.}

\begin{equation}\label{ra_mask}
    \mathcal{M}_{RA}^i=\epsilon[1\ominus Softmax\{\mathcal{U}(S_{i+1}(j)\}],
\end{equation}

\subsection{Boundary Attention}\label{boundary}

\begin{figure*}
    \centering
    \includegraphics[scale=0.45]{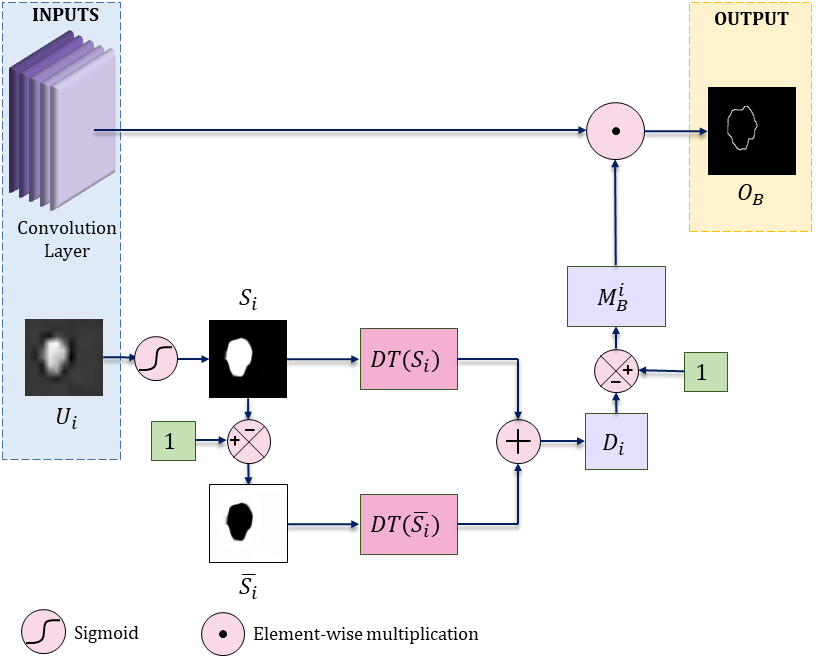}
    \caption{Architecture of the BA module used in the MFSNet model.\\ $U_i$: Global map output from PPD module; $S_i$: Segmentation Map; $\bar{S_i}$: Inverted Segmentation Map; $DT(x)$: Distance Transform; $\mathcal{M}_B^i$: $i^{th}$ level boundary mask; $O_B$: Boundary Attention output}
    \label{ba_fig}
\end{figure*}

Edge information can guide the task of feature extraction for segmentation by providing helpful supervision with fine-grained boundary constraints as shown in \cite{zhao2019egnet}. Hence, being inspired by the Edge Guidance Module (EGM), proposed by \cite{zhang2019net}, we have used a BA module along with the parallel RA branches for extracting accurate boundary information. Based on the fact that only low-level features contain substantial edge information, we have fed the shallow feature $F_2$ from the encoder network to the BA module as shown in \autoref{ba_fig}. \textcolor{black}{The BA module helps the network capture important boundary information, which is complementary to the amplified reverse class response for the regions of shared semantic information extracted by the RA module. This additional edge information acts as a helpful signal to confusing segment regions near the lesion boundaries.} The $i^{th}$ level feature $F_i$ from the encoder, when fed to the BA module, produces an output $O_B$, given by \autoref{ba}, where $\odot$ is the element-wise multiplication of feature $F_i$ and the $i^{th}$ level boundary mask $\mathcal{M}_B^i$, which is obtained by formulating the binary segmentation map $S_i$ given by \autoref{binary}, where $j$ is the pixel position index, $U_i$ denotes the $i^{th}$ level upsampled prediction.

\begin{equation}\label{ba}
    O_B(F_i)=\mathcal{M}_B^i\odot F_i,
\end{equation}

\begin{equation}\label{binary}
    S_i(j)=\begin{cases}
    1, & \text{if }\sigma [U_i(j)]>0.5\\
    0, & \text{otherwise}
    \end{cases},
\end{equation}

The value of $i$ is set to 2, 3, i.e., we only consider the second and third level features from the CNN to feed into the BA module. $\sigma$ is the softmax activation function given by the \autoref{softmax}.

\begin{equation}\label{softmax}
    \sigma(x_i)=\frac{\exp x_m}{\sum\limits_{n}\exp x_n}
\end{equation}

Next, distance transformation \cite{fabbri20082d} is applied over $S_i$ to fill each pixel position of the melanoma region with the distance to the melanoma boundary. Conversely, the distances of the pixels of non-melanoma regions can be obtained by simply transposing $S_i$ followed by distance transformation. The overall distance map is produced by normalizing and summing up these two distance maps as given by \autoref{sum}, where $\overline{S_i}$ is the transpose of segmentation map $S_i$ which can be obtained as $\overline{S_i}=1-S_i$.

\begin{equation}\label{sum}
    D_i=\frac{DT(S_i)}{\max_j DT[S_i(j)]}+\frac{DT(\overline{S_i})}{\max_j DT[\overline{S_i(j)}]},
\end{equation}

 In \autoref{sum}, $D_i$ has values equal to 0 and 1 at the melanoma boundary and the farthest point from the boundary, respectively. Here, we define the $i^{th}$ level boundary mask $\mathcal{M_B}^i$ as

\begin{equation}
    \mathcal{M_B}^i=1-D_i
\end{equation}

Finally, we calculate the boundary map $G_B$ from the ground truth using its gradient, which is constrained by the BCE loss to measure the dissimilarity between the produced boundary map $O_B$ with the actual boundary map $G_B$ given by \autoref{bce}.
\begin{equation}\label{bce}
    \mathcal{L_B}=-\sum\limits_{j}[G_B\log(O_B)+(1-G_B)\log(1-O_B)]
\end{equation}
The overall architecture of the BA module is shown in \autoref{ba_fig}.

\subsection{Partial Parallel Decoder module}\label{PPD}
\begin{figure*}
    \centering
    \resizebox{\textwidth}{!}{%
    \includegraphics[scale=0.6]{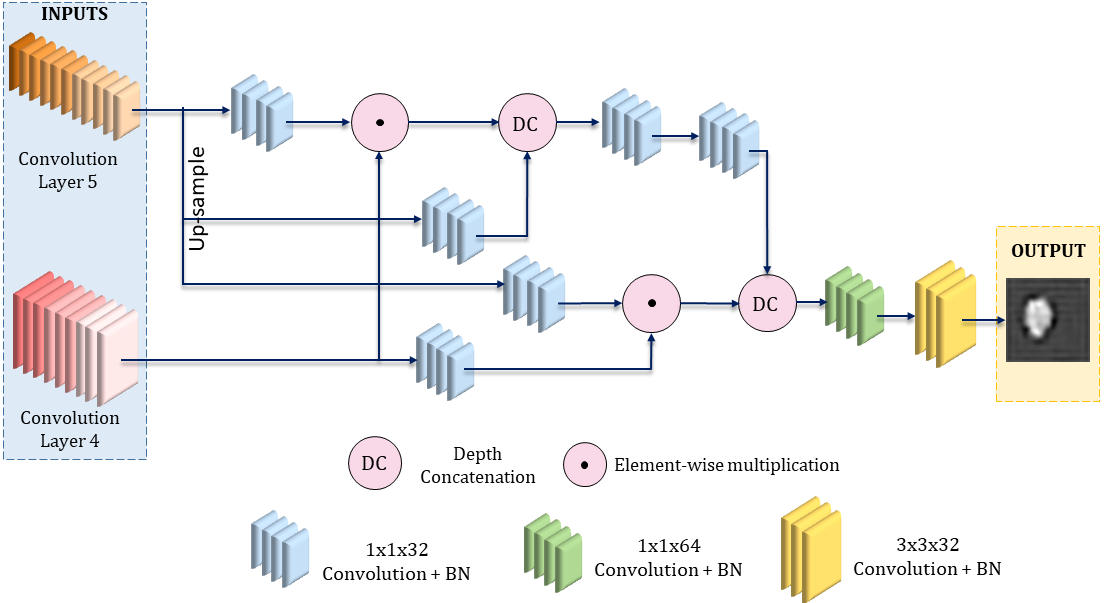}
    }
    \caption{Architecture of the Partial Parallel Decoder module used in the proposed framework. The convolution layers 4 and 5 denote the 4th and fifth layer respectively of the Res2Net CNN backbone used in MFSNet.}
    \label{ppdfig}
\end{figure*}

As suggested by \cite{wu2019cascaded}, low-level features contribute very little toward the final prediction map with a massive requirement of computation due to their high spatial resolution. However, in literature, most of the existing models like \cite{zhou2018unet++,gu2019net} are designed to aggregate both high and low-level semantics, leading to unnecessary wastage of resources and inefficient segmentation map. To mitigate this problem, we have used a PPD module to capture the global context information, being inspired from the Receptive Field Block (RFB) module by \cite{liu2018receptive}. 

Specifically, we have used the first five convolution layers of the Res2Net \cite{gao2019res2net}, among which the first three layers are considered as the low-level features and are discarded for the decoder module. To accelerate the feature propagation, we add a series of convolution and batch normalization operations as shown in \autoref{ppdfig}. Short connections are added in the PPD module, similar to the original RFB module. After obtaining different discriminating features from different layers, we finally multiply them to reduce the gap between multiple feature levels. Thus, the PPD module produces a global segmentation map $O_S$ through a series of element-wise multiplication and concatenation operations, serving as the global guidance of the parallel RA branches. \textcolor{black}{Proper downsampling and upsampling operations are performed throughout, whenever required, to match the feature dimensions before concatenation.} Finally, the generated segmentation map is of a similar dimension as that of the input of the MFSNet.

\subsection{Deep Supervision}\label{loss}
To supervise the segmentation performance, we have used a hybrid loss function in this research. For the BA module, we have used the standard BCE loss function, shown in \autoref{bce}. However, for the supervision of segmentation, we have used a mixing loss function for effective global and local supervision to enhance both image-level and pixel-level segmentation, respectively. The proposed loss function involves the weighted BCE loss function $\mathcal{L}_{wBCE}$ and weighted IoU loss function $\mathcal{L}_wIoU$, given by \autoref{hybridloss}, where $\delta$ is the weight, set to 0.9 in our case experimentally.

\begin{equation}\label{hybridloss}
    \mathcal{L}_S=\delta\mathcal{L}_{wBCE}+(1-\delta)\mathcal{L}_{wIoU},
\end{equation}

\begin{figure}
    \centering
    \resizebox{\textwidth}{!}{
    \includegraphics{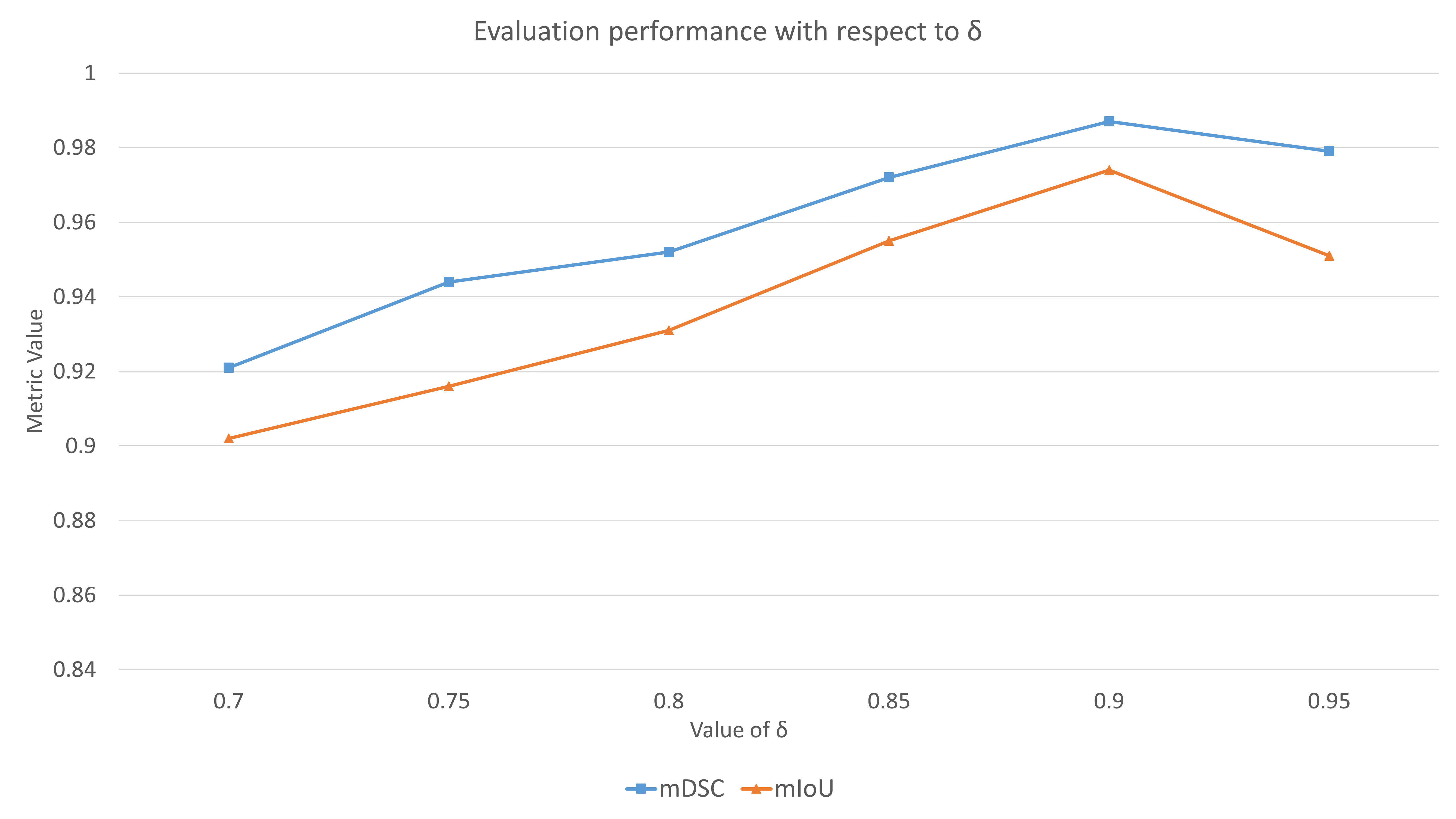}
    }
    \caption{Experimental analysis of mean DSC and mean IoU on ISIC2017 dataset against different $\delta$ values, that defined the weights of different components in the proposed loss function (\autoref{hybridloss}).}
    \label{fig:hyperparameter}
\end{figure}

The experimental analysis is shown in \autoref{fig:hyperparameter}. The $\mathcal{L}_{wIoU}$ and $\mathcal{L}_{wBCE}$ are effective to increase the weights of the hard pixels rather than giving equal weights to each pixel like the standard IoU loss and BCE loss functions. 

The side outputs from the CNN are upsampled to form segmentation map $O^{UP}_i;\:{i=4,5}$ of the same size of the ground truth $G$. Thus, the overall loss function is extended to \autoref{finalloss}.
\begin{equation}\label{finalloss}
    \mathcal{L}=\mathcal{L}_S(G,O_S)+\sum\limits_{i=2,3}\mathcal{L}_B(G_B,O_{B,i})+\sum\limits_{i=4,5}\mathcal{L}_S(G,O_i^{UP})
\end{equation}

\section{Results and Discussion}\label{results}
This section evaluates the proposed framework on three publicly available datasets of skin lesion segmentation, using 5-fold cross-validation. We discuss the significance of the obtained results and compare the model with other state-of-the-art models to justify the superiority of the proposed model.

\subsection{Dataset Description}
Three dermatology datasets have been used in the current research to evaluate the performance of MFSNet: 
\begin{enumerate}
    \item $PH^2$ dataset by \cite{mendoncca2013ph} consisting of 200 images.
    \item ISIC 2017 dataset by \cite{codella2018skin} consisting of 2379 images.
    \item HAM10000 dataset by \cite{tschandl2018ham10000} consisting of 10015 images.
\end{enumerate}

\subsection{Evaluation Metrics}\label{experiment}
To evaluate the performance of the proposed model on the supervised skin lesion segmentation problem, we use five popularly used metrics which are described as follows:

\begin{enumerate}
    
    \item Dice Similarity Coefficient (DSC): It is a spatial overlap metric which is computed as in \autoref{meanDSC} for predicted image $S$ and ground truth $G$. 
    \begin{equation}\label{meanDSC}
        DSC(S,G)=\frac{2\times |S\cap G|}{|S|+|G|}
    \end{equation}

    \item Intersection over Union (IoU): IoU, also known as Jaccard Index (JI), measures segmentation accuracy by computing the ratio of the intersection of objects and their union when projected on the same plane. Mathematically it is expressed as in \autoref{meanIoU}, where $S$ is the predicted segmentation mask, and $G$ is the original ground truth mask of the image.
    \begin{equation}\label{meanIoU}
        IoU(S,G)=\frac{|S\cap G|}{|S\cup G|}
    \end{equation}
    
    \item F-Measure (FM): F-Measure is a standard metric that evaluates the harmonic mean of the pixel-wise precision and recall and is mathematically expressed as in \autoref{wfm}.
    \begin{equation}\label{wfm}
        FM=\frac{2\times Precision\times Recall}{Recall+Precision}
    \end{equation}
    
    \item  Sensitivity (Sen): It characterizes the percentage of pixels of the object that are accurately classified as the object class and it is computed by \autoref{sensi}.
    \begin{equation}\label{sensi}
        Sen(S,G)=\frac{|S\cap G|}{|G|}
    \end{equation}
    
    \item Specificity (Spe): It characterizes the percentage of pixels of the background class that are accurately classified as the background, and it is computed using \autoref{speci}.
    \begin{equation}\label{speci}
        Spe(S,G)=\frac{|(1-S)\cap(1-G)|}{|1-G|}
    \end{equation}
    
\end{enumerate}
For all the mentioned evaluation metrics, the mean value over all the test images has been reported in this study for evaluation denoted by $mIoU$, $mDSC$, $mFM$, $mSen$ and $mSpe$.

\subsection{Implementation}

\begin{table*}[]
\centering
\caption{Results obtained by MFSNet on the three datasets using 5-fold cross-validation.}
\label{res}
\resizebox{\textwidth}{!}{%
\begin{tabular}{|c|c|c|c|c|c|c|}
\hline
\textbf{Dataset} & \textbf{Fold} & \textbf{mDSC} & \textbf{mIoU} & \textbf{mFM} & \textbf{mSen} & \textbf{mSpe} \\ \hline
\multirow{6}{*}{$PH^2$}      & 1                & 0.955                & 0.917                & 0.947                & 1.000                & 1.000                \\ \cline{2-7} 
                          & 2                & 0.956                & 0.918                & 0.941                & 0.991                & 0.986                \\ \cline{2-7} 
                          & 3                & 0.951                & 0.915                & 0.945                & 0.995                & 0.999                \\ \cline{2-7} 
                          & 4                & 0.949                & 0.920                & 0.943                & 1.000                & 1.000                \\ \cline{2-7} 
                          & 5                & 0.958                & 0.899                & 0.941                & 0.989                & 0.999                \\ \cline{2-7} 
                          & \textbf{Average} & \textbf{0.954±0.003} & \textbf{0.914±0.008} & \textbf{0.944±0.002} & \textbf{0.995±0.004} & \textbf{0.997±0.002} \\ \hline
\multirow{6}{*}{ISIC 2017} & 1                & 0.991                & 0.976                & 0.989                & 1.000                & 1.000                \\ \cline{2-7} 
                          & 2                & 0.985                & 0.971                & 0.980                & 1.000                & 1.000                \\ \cline{2-7} 
                          & 3                & 0.983                & 0.967                & 0.980                & 0.998                & 0.999                \\ \cline{2-7} 
                          & 4                & 0.986                & 0.980                & 0.991                & 0.999                & 0.999                \\ \cline{2-7} 
                          & 5                & 0.990                & 0.975                & 0.989                & 0.997                & 0.998                \\ \cline{2-7} 
                          & \textbf{Average} & \textbf{0.987±0.003} & \textbf{0.974±0.004} & \textbf{0.986±0.005} & \textbf{0.999±0.001} & \textbf{0.999±0.001} \\ \hline
\multirow{6}{*}{HAM10000} & 1                & 0.911                & 0.910                & 0.905                & 1.000                & 0.999                \\ \cline{2-7} 
                          & 2                & 0.900                & 0.901                & 0.899                & 0.997                & 0.998                \\ \cline{2-7} 
                          & 3                & 0.905                & 0.903                & 0.906                & 0.999                & 1.000                \\ \cline{2-7} 
                          & 4                & 0.904                & 0.894                & 0.892                & 1.000                & 1.000                \\ \cline{2-7} 
                          & 5                & 0.910                & 0.900                & 0.914                & 0.998                & 0.998                \\ \cline{2-7} 
                          & \textbf{Average} & \textbf{0.906±0.004} & \textbf{0.902±0.005} & \textbf{0.903±0.007} & \textbf{0.999±0.001} & \textbf{0.999±0.001} \\ \hline
\end{tabular}
}
\end{table*}

\begin{table*}[]
\centering
\caption{Comparison of the results obtained with the MFSNet model on the three datasets with and without image preprocessing}
\label{tab:comp_preprocess}
\resizebox{0.8\textwidth}{!}{%
\begin{tabular}{|
c |
c |
c |
c |
c |
c |}
\hline
\textbf{Dataset}                                   & \textbf{Preprocessing} & \textbf{mDSC}   & \textbf{mIoU}   & \textbf{mSen} & \textbf{mSpe} \\ \hline
                           & NO                     & 0.931          & 0.895          & 0.978                & 0.978                \\ \cline{2-6} 
\multirow{-2}{*}{PH2}      & \textbf{YES}           & \textbf{0.954} & \textbf{0.914} & \textbf{0.995}       & \textbf{0.997}       \\ \hline
                           & NO                     & 0.963          & 0.942          & 0.969                & 0.970                \\ \cline{2-6} 
\multirow{-2}{*}{ISIC2017} & \textbf{YES}           & \textbf{0.987} & \textbf{0.974} & \textbf{0.999}       & \textbf{0.999}       \\ \hline
                           & NO                     & 0.872          & 0.869          & 0.954                & 0.961                \\ \cline{2-6} 
\multirow{-2}{*}{HAM10000} & \textbf{YES}           & \textbf{0.906} & \textbf{0.902} & \textbf{0.999}       & \textbf{0.999}       \\ \hline
\end{tabular}%
}
\end{table*}

The proposed MFSNet is implemented in PyTorch and is accelerated using an NVIDIA Tesla K80 GPU. \autoref{res} shows the results obtained by MFSNet on the three publicly available datasets using 5-fold cross-validation. The high values of $DSC$ and $IoU$ suggest that the segmentation is reasonably accurate. In contrast, the high Sensitivity and Specificity values suggest the maintenance of structural coherence between the segmented mask and the available ground-truth mask. Further, to evaluate the importance of image preprocessing (artifact removal) in this research, we evaluate and compare the performance of the MFSNet model with the raw images and the preprocessed images. The results of these experiments are presented in \autoref{tab:comp_preprocess}.

\subsection{Ablation Study}\label{ablation}
We have experimented by removing different components from the proposed model to justify their impact on the overall performance. We have performed an ablation study of RA, PPD, BA modules and their different orientations concerning the convolution layers of the backbone Res2Net model to assert the importance of the proposed configuration used in the MFSNet architecture.

\subsubsection{Orientation of BA and RA}
We have experimented with different combinations and orientations of BA and RA branches to explore the best possible combinations for boosting performance. \autoref{bara} shows the results on the $PH^2$ dataset, where we have used RA and BA modules at different levels of feature extraction. Comparing instances 1 and 4 from the table shows that the performance can be boosted if we use BA at the $Conv2$ layer instead of RA. This behavior can be justified because the shallow layers of the CNN can extract features rich in boundary information. Hence adding BA there will provide additional edge guidance to the model. Similar conclusions can be drawn by comparing instances 1 and 2. Again, comparing instances 2 and 3, we can observe experimentally that removing the BA module from the $Conv1$ layer does not decrease the segmentation performance significantly but effectively reduces computation of an additional BA module. We have slightly better performance in instance three than in instance 5, establishing the importance of the RA module at the $Conv4$ layer. Based on these observations, we have finalized the orientations of different RA and BA blocks to optimize the segmentation performance and add their clinical importance.

\begin{table*}[]
\centering
\caption{Comparison of quantitative results obtained from different orientations of RA and BA blocks in the proposed MFSNet model on the $PH^2$ dataset. The highlighted row indicates the orientations and results of our proposed model.}
\label{bara}
\resizebox{\textwidth}{!}{%
\begin{tabular}{|c|c|c|c|c|c|c|c|c|c|c|}
\hline
                           & \multicolumn{5}{c|}{\textbf{Combinations}}     & \multicolumn{5}{c|}{\textbf{Average Result (on 5 fold)}}                               \\ \cline{2-11} 
\multirow{-2}{*}{\textbf{Instance}} & \textbf{Conv1} & \textbf{Conv2} & \textbf{Conv3} & \textbf{Conv4} & \textbf{Conv5} & \textbf{mDSC}    & \textbf{mIoU}    & \textbf{mFM}   & \textbf{mSen} & \textbf{mSpe} \\ \hline
1                          & BA    & BA    & BA    & RA    & RA    & 0.944±0.002 & 0.897±0.003 & 0.928±0.004 & 0.984±0.008      & 0.989±0.004      \\ \hline
2                          & BA    & BA    & RA    & RA    & RA    & 0.926±.006  & 0.872±0.002 & 0.902±0.006 & 0.971±0.004      & 0.969±0.006      \\ \hline
3                          & -     & BA    & RA    & RA    & RA    & 0.930±0.004 & 0.876±0.003 & .911±0.007  & 0.979±0.005      & 0.972±0.006      \\ \hline
4                          & BA    & RA    & BA    & RA    & RA    & 0.926±0.004 & 0.876±0.002 & 0.916±0.005 & 0.966±0.004      & 0.963±0.002      \\ \hline
5                          & -     & BA    & RA    & BA    & RA    & 0.926±0.006 & 0.871±0.005 & 0.902±0.003 & 0.978±0.006      & 0.970±0.004      \\ \hline
 \textbf{Proposed} &
    \textbf{-} &
    \textbf{BA} &
    \textbf{BA} &
    \textbf{RA} &
    \textbf{RA} &
    \textbf{0.954±0.003} &
    \textbf{0.914±0.008} &
    \textbf{0.944±0.002} &
    \textbf{0.995±0.004} &
    \textbf{0.997±0.002} \\ \hline
\end{tabular}%
}
\end{table*}

\begin{table*}[]
\centering
\caption{Results of the ablation study considering various components of the MFSNet model on the $PH^2$ dataset. Best results are highlighted.}
\label{abtab}
\resizebox{\textwidth}{!}{
\begin{tabular}{|c |c |c |c |c |c |}
\hline
\textbf{Architecture}                 & \textbf{mDSC}        & \textbf{mIoU}        & \textbf{mFM}         & \textbf{mSen}        & \textbf{mSpe}        \\ \hline
Res2Net        & 0.794±0.006 & 0.758±0.008 & 0.761±0.005 & 0.816±0.009 & 0.821±0.008 \\ \hline
Res2Net+PPD    & 0.877±0.005 & 0.852±0.004 & 0.873±0.003 & 0.915±0.006 & 0.906±0.004 \\ \hline
Res2Net+BA     & 0.843±0.003 & 0.820±0.006 & 0.871±0.004 & 0.911±0.007 & 0.904±0.004 \\ \hline
Res2Net+RA     & 0.842±0.002 & 0.834±0.005 & 0.866±0.006 & 0.909±0.006 & 0.929±0.004 \\ \hline
Res2Net+BA+RA  & 0.906±0.003 & 0.861±0.004 & 0.894±0.006 & 0.947±0.004 & 0.936±0.007 \\ \hline
Res2Net+RA+PPD & 0.927±0.003 & 0.895±0.007 & 0.912±0.005 & 0.963±0.007 & 0.959±0.006 \\ \hline
\textbf{Res2Net+BA+RA+PPD (Proposed)} & \textbf{0.954±0.003} & \textbf{0.914±0.008} & \textbf{0.944±0.002} & \textbf{0.995±0.004} & \textbf{0.997±0.002} \\ \hline
\end{tabular}
}
\end{table*}

\subsubsection{Importance of BA}
In this work, we have also performed an ablation study to investigate the importance of the proposed BA module in the overall model. Row 3 in \autoref{abtab} shows the performance of the proposed architecture has improved by a considerable margin in terms of significant evaluation metrics by using the BA module along with the $Res2Net$ backbone as compared to the backbone only in row 1. Besides, using BA along with RA boosts the model performance as compared to only the RA module, shown in row 4 and row 5 of \autoref{abtab}, leading to the conclusion that BA has an essential contribution towards achieving better segmentation outcome. \cite{zhang2019net} also exemplified that optimal edge guidance can boost the segmentation performance significantly, justifying the results obtained in our experiment.

\subsubsection{Importance of RA}
Row 4 in \autoref{abtab} shows that RA is another essential component of the proposed module as removing it may reduce the DSC, IoU, and other evaluation results significantly. The optimal combination of BA and RA modules (shown in \autoref{bara}) is another essential feature of the proposed MFSNet, where the addition of RA has boosted the model performance as compared to the mere BA module, shown in row 5 of \autoref{abtab}.

\subsubsection{Importance of PPD}
PPD is another vital component of our proposed method, as removal of this can affect the model performance as shown in \autoref{abtab}. We can observe from row 2 of \autoref{abtab} that adding PPD to the baseline model can increase the performance, unparalleled to the contribution of RA and BA. Again, combining it with the RA module, as shown in row 6, can produce an almost similar performance to that of the proposed architecture. The improvements establish that PPD, combined with RA, is the prime component of the proposed MFSNet.

\subsection{Comparison to State-of-the-art}\label{SOTA}

\begin{table}[]
\centering
\caption{Comparison of the proposed MFSNet model to state-of-the-art models on the three publicly available datasets used in this study. (Total training time is calculated on implementation using NVIDIA Tesla K80 GPU)}
\label{comp_lit}
\resizebox{\textwidth}{!}{
\begin{tabular}{|c|c|c|c|c|c|c|}
\hline
\textbf{Dataset}                     & \textbf{Model}           & \textbf{mDSC}  & \textbf{mIoU}  & \textbf{mSen}  & \textbf{mSpe}  & \textbf{Training time} \\ \hline
 & Double   U-Net \cite{jha2020doubleu}          & 0.907 & 0.899 & 0.945  & 0.966 & 1hr 2min12sec            \\ \cline{2-7} 
 & U-Net \cite{ronneberger2015u}                 & 0.876 & 0.780 & 0.816  & 0.978 & 30min 54 sec             \\ \cline{2-7} 
 & SegNet \cite{badrinarayanan2017segnet}        & 0.894 & 0.808 & 0.865  & 0.966 & 58min 21 sec             \\ \cline{2-7} 
 & Goyal et   al. \cite{goyal2019skin}           & 0.907 & 0.839 & 0.932  & 0.929 & -                        \\ \cline{2-7} 
 & Hasan et al. \cite{hasan2020dsnet}            & -     & 0.870 & 0..929 & 0.969 & 35min 08sec              \\ \cline{2-7} 
 & Al et al. \cite{al2018skin}                   & 0.918 & 0.848 & 0.937  & 0.957 & -                        \\ \cline{2-7} 
 & Ozturk et al. \cite{ozturk2020skin}           & 0.930 & 0.871 & 0.969  & 0.953 & -                        \\ \cline{2-7} 
 & Xie et al. \cite{xie2020skin}                 & 0.919 & 0.857 & 0.963  & 0.942 & -                        \\ \cline{2-7} 
 & Unver et al. \cite{unver2019skin}             & 0.881 & 0.795 & 0.836  & 0.940 & -                        \\ \cline{2-7} 
 & Yuan et al. \cite{yuan2017automatic}          & 0.915 & -     & -      & -     & -                        \\ \cline{2-7} 
 & Bi et al. \cite{bi2017dermoscopic}            & 0.907 & 0.840 & 0.949  & 0.940 & -                        \\ \cline{2-7} 
 & Bi et al. \cite{bi2019step}                   & 0.921 & 0.859 & 0.962  & 0.945 & -                        \\ \cline{2-7} 
\multirow{-13}{*}{\textbf{$PH^2$}}   & \textbf{Proposed MFSNet} & \textbf{0.954} & \textbf{0.914} & \textbf{0.995} & \textbf{0.997} & \textbf{46min 37sec}         \\ \hline
 & Double   U-Net \cite{jha2020doubleu}          & 0.913 & 0.918 & 0.963  & 0.974 & 4hr 18min 07sec          \\ \cline{2-7} 
 & U-Net \cite{ronneberger2015u}                 & 0.778 & 0.683 & 0.812  & 0.805 & \textbf{3hr 22min 44sec} \\ \cline{2-7} 
 & SegNet \cite{badrinarayanan2017segnet}        & 0.821 & 0.696 & 0.801  & 0.954 & 4hr 04min 17sec          \\ \cline{2-7} 
 & Tschandl et al. \cite{tschandl2019domain}     & 0.853 & 0.770 & -      & -     & -                        \\ \cline{2-7} 
 & Navarro et al. \cite{navarro2018accurate}     & 0.938 & 0.846 & -      & -     & -                        \\ \cline{2-7} 
 & Saha et al. \cite{saha2020leveraging}         & 0.855 & 0.772 & 0.824  & 0.981 & -                        \\ \cline{2-7} 
 & Goyel et al. \cite{goyal2019skin}             & 0.793 & 0.871 & 0.899  & 0.950 & -                        \\ \cline{2-7} 
 & Hasan et al. \cite{hasan2020dsnet}            & -     & 0.775 & 0.875  & 0.955 & 3hr 37min 17sec          \\ \cline{2-7} 
 & Al et al. \cite{al2018skin}                   & 0.871 & 0.771 & 0.854  & 0.967 & -                        \\ \cline{2-7} 
 & Ozturk et al. \cite{ozturk2020skin}           & 0.886 & 0.783 & 0.854  & 0.981 & -                        \\ \cline{2-7} 
 & Xie et al. \cite{xie2020skin}                 & 0.862 & 0.783 & 0.870  & 0.964 & -                        \\ \cline{2-7} 
 & Unver et al. \cite{unver2019skin}             & 0.843 & 0.748 & 0.908  & 0.927 & -                        \\ \cline{2-7} 
\multirow{-13}{*}{\textbf{ISIC2017}} & \textbf{Proposed MFSNet} & \textbf{0.987} & \textbf{0.974} & \textbf{0.999} & \textbf{0.999} & 3hr 51min 20sec              \\ \hline
 & Double   U-Net \cite{jha2020doubleu}          & 0.843 & 0.812 & 0.861  & 0.845 & 11hr 21min 53sec         \\ \cline{2-7} 
 & U-Net \cite{ronneberger2015u}                 & 0.781 & 0.774 & 0.799  & 0.802 & 9hr 05min 31sec          \\ \cline{2-7} 
 & SegNet \cite{badrinarayanan2017segnet}        & 0.816 & 0.821 & 0.867  & 0.854 & 11hr 04min 10sec         \\ \cline{2-7} 
 & Saha et al. \cite{saha2020leveraging}         & 0.891 & 0.819 & 0.824  & 0.981 & -                        \\ \cline{2-7} 
 & Abraham et al. \cite{abraham2019novel}        & 0.856 & -     & -      & -     & -                        \\ \cline{2-7} 
 & Shahin et al. \cite{shahin2019deep}           & 0.903 & 0.837 & 0.902  & 0.974 & -                        \\ \cline{2-7} 
 & Bissoto et al. \cite{bissoto2018deep}         & 0.873 & 0.792 & 0.934  & 0.936 & -                        \\ \cline{2-7} 
 & Ibtehaz et al. \cite{ibtehaz2020multiresunet} & -     & 0.803 & -      & -     & -                        \\ \cline{2-7} 
\multirow{-9}{*}{\textbf{HAM10000}}  & \textbf{Proposed MFSNet} & \textbf{0.906} & \textbf{0.902} & \textbf{0.999} & \textbf{0.999} & \textbf{9hr 41min 34sec}     \\ \hline
\end{tabular}
}
\end{table}

\begin{figure*}
    \centering
    \resizebox{\textwidth}{!}{%
    \includegraphics{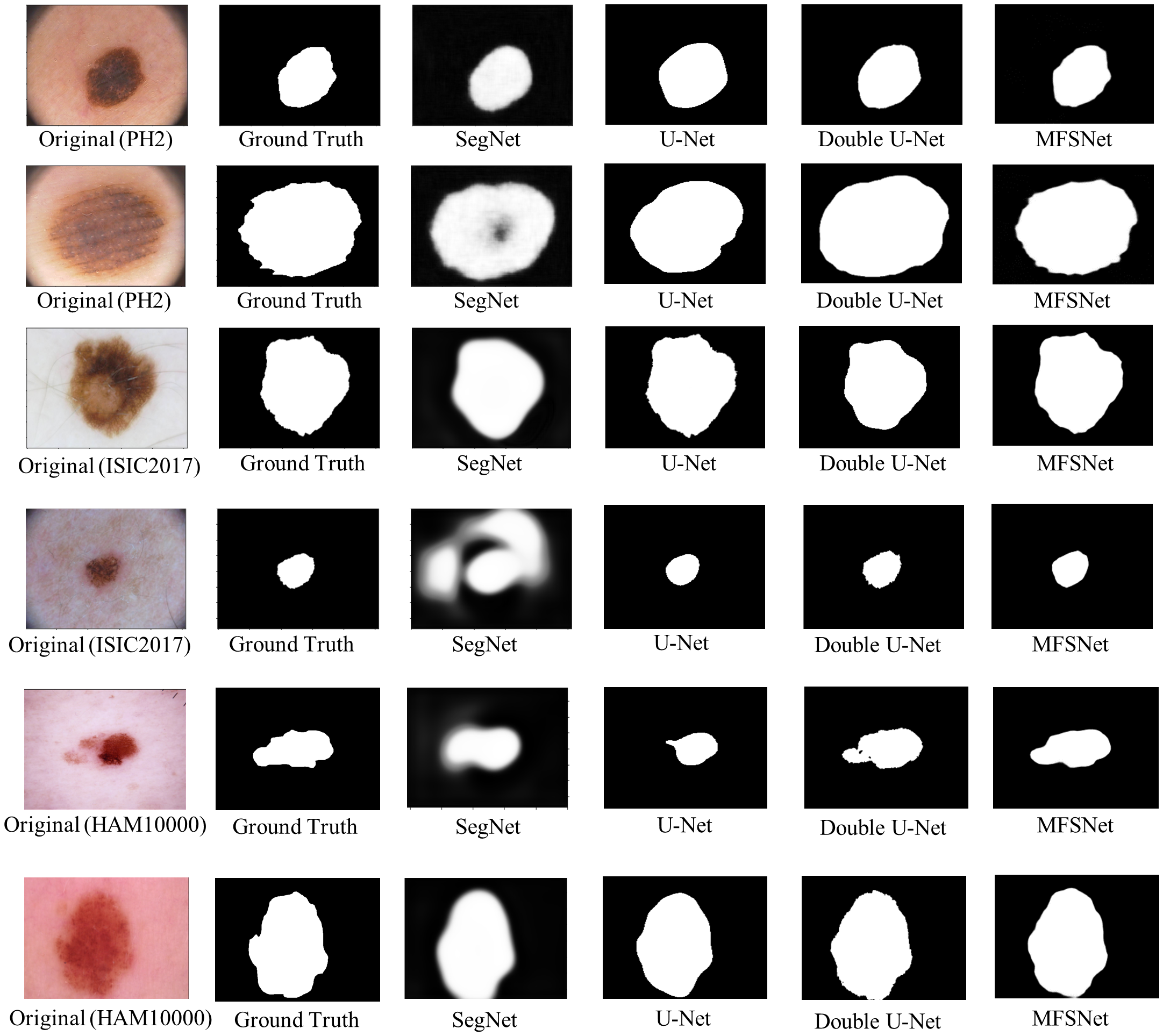}
    }
    \caption{A few instances of segmentation masks obtained by some standard models in literature compared to the proposed MFSNet model.}
    \label{comp_std_fig}
\end{figure*}

\autoref{comp_lit} compares the proposed method to several state-of-the-art methods on the three datasets used. The proposed MFSNet performs significantly better than the said methods and can be justified as a reliable framework for skin lesion segmentation. To further prove the superiority of the MFSNet framework, we use some popular segmentation models prevalent in literature for comparison: U-Net \cite{ronneberger2015u}, SegNet \cite{badrinarayanan2017segnet} and Double U-Net \cite{jha2020doubleu}, the results of which are also compared in \autoref{comp_lit}. Some visual results of the predicted segmented masks by these models and the proposed MFSNet are shown in \autoref{comp_std_fig}. From the visual results, it can be seen that SegNet consistently produces unsatisfactory results. U-Net can segment most images well, but it fails to perform well for relatively challenging images. Double U-Net performs closest to the MFSNet. However evidently MFSNet outperforms all these models as justified from both \autoref{comp_lit} and \autoref{comp_std_fig}.

We have also compared the computational cost of MFSNet in terms of the total execution (training) time with existing methods, as shown in \autoref{comp_lit}. It is clear from the table that our proposed method is computationally efficient compared to several state-of-the-art methods like SegNet, DoubleUNet, etc. However, we could not calculate the execution time for all the methods in the literature compared in this study due to the unavailability of open-source implementations.

The improvement in the values of the evaluation metrics by the MFSNet model as compared to state-of-the-art methods in the literature can significantly impact the diagnosis process. Higher values of IoU, DSC, etc., indicate a more accurate skin lesion segmentation while preserving structural similarity. Thus, when the segmented lesions are used for further diagnosis, more robust and informative features can be extracted for the automatic classification of the lesions into benign and malignant classes as stated by Mahbod et al. \cite{mahbod2020effects}. This reduces the chances of faulty diagnosis and helps control skin cancer early and more effectively.

\section{Conclusion and Future Work}\label{conclusions}
The emergence of CAD systems has facilitated several seemingly daunting tasks, like the segmentation of skin lesions. Skin cancer affects a large population worldwide, and hence its early detection is essential for eradicating cancer. Localization of tumors and lesion segmentation poses a challenge since an esoteric group of clinicians can only perform manual segmentation, and it is also a time-demanding task. To bolster the efforts of the medical practitioners, in this research, we develop a fully automated framework for accurate skin lesion segmentation from raw dermoscopy images. The proposed framework uses multi-scaled maps using a PPD module and two RA and BA modules to produce the final segmentation mask. The use of the multi-focus-based approach helps determine the overall lesion structure from the coarse map. The use of the finer maps helps in determining more refined edges, leading to increased segmentation accuracy. Upon evaluating the proposed MFSNet model on three publicly available datasets of varied sizes, the proposed method displayed robust performance, outperforming the state-of-the-art methods on the respective datasets.

In the future, we may extend the segmentation model to other domains like brain MRIs, lung CT scans, etc. Also, we might incorporate semi-supervision for the segmentation, similar to \cite{li2021single} to extend the models to unlabelled datasets.

\section*{Acknowledgements}
The authors would like to thank the Centre for Microprocessor Applications for Training, Education and Research (CMATER) laboratory of the Computer Science and Engineering Department, Jadavpur University, Kolkata, India, for providing the infrastructural support.

\section*{Conflict of interest}
All the authors declare that there is no conflict of interest.

\bibliography{MFSNet_unhighlighted}

\begin{thebibliography}{57}
\expandafter\ifx\csname natexlab\endcsname\relax\def\natexlab#1{#1}\fi
\providecommand{\url}[1]{\texttt{#1}}
\providecommand{\href}[2]{#2}
\providecommand{\path}[1]{#1}
\providecommand{\DOIprefix}{doi:}
\providecommand{\ArXivprefix}{arXiv:}
\providecommand{\URLprefix}{URL: }
\providecommand{\Pubmedprefix}{pmid:}
\providecommand{\doi}[1]{\href{http://dx.doi.org/#1}{\path{#1}}}
\providecommand{\Pubmed}[1]{\href{pmid:#1}{\path{#1}}}
\providecommand{\bibinfo}[2]{#2}
\ifx\xfnm\relax \def\xfnm[#1]{\unskip,\space#1}\fi
\bibitem[{Abraham \& Khan(2019)}]{abraham2019novel}
\bibinfo{author}{Abraham, N.}, \& \bibinfo{author}{Khan, N.~M.}
  (\bibinfo{year}{2019}).
\newblock \bibinfo{title}{A novel focal tversky loss function with improved
  attention u-net for lesion segmentation}.
\newblock In {\it \bibinfo{booktitle}{2019 IEEE 16th International Symposium on
  Biomedical Imaging (ISBI 2019)}\/} (pp. \bibinfo{pages}{683--687}).
\newblock \bibinfo{organization}{IEEE}.
\bibitem[{Al-Masni et~al.(2018)Al-Masni, Al-Antari, Choi, Han \&
  Kim}]{al2018skin}
\bibinfo{author}{Al-Masni, M.~A.}, \bibinfo{author}{Al-Antari, M.~A.},
  \bibinfo{author}{Choi, M.-T.}, \bibinfo{author}{Han, S.-M.}, \&
  \bibinfo{author}{Kim, T.-S.} (\bibinfo{year}{2018}).
\newblock \bibinfo{title}{Skin lesion segmentation in dermoscopy images via
  deep full resolution convolutional networks}.
\newblock {\it \bibinfo{journal}{Computer methods and programs in
  biomedicine}\/},  {\it \bibinfo{volume}{162}\/}, \bibinfo{pages}{221--231}.
\bibitem[{Aljanabi et~al.(2018)Aljanabi, {\"O}zok, Rahebi \&
  Abdullah}]{aljanabi2018skin}
\bibinfo{author}{Aljanabi, M.}, \bibinfo{author}{{\"O}zok, Y.~E.},
  \bibinfo{author}{Rahebi, J.}, \& \bibinfo{author}{Abdullah, A.~S.}
  (\bibinfo{year}{2018}).
\newblock \bibinfo{title}{Skin lesion segmentation method for dermoscopy images
  using artificial bee colony algorithm}.
\newblock {\it \bibinfo{journal}{Symmetry}\/},  {\it \bibinfo{volume}{10}\/},
  \bibinfo{pages}{347}.
\bibitem[{Attia et~al.(2017)Attia, Hossny, Nahavandi \&
  Yazdabadi}]{attia2017skin}
\bibinfo{author}{Attia, M.}, \bibinfo{author}{Hossny, M.},
  \bibinfo{author}{Nahavandi, S.}, \& \bibinfo{author}{Yazdabadi, A.}
  (\bibinfo{year}{2017}).
\newblock \bibinfo{title}{Skin melanoma segmentation using recurrent and
  convolutional neural networks}.
\newblock In {\it \bibinfo{booktitle}{2017 IEEE 14th International Symposium on
  Biomedical Imaging (ISBI 2017)}\/} (pp. \bibinfo{pages}{292--296}).
\newblock \bibinfo{organization}{IEEE}.
\bibitem[{Badrinarayanan et~al.(2017)Badrinarayanan, Kendall \&
  Cipolla}]{badrinarayanan2017segnet}
\bibinfo{author}{Badrinarayanan, V.}, \bibinfo{author}{Kendall, A.}, \&
  \bibinfo{author}{Cipolla, R.} (\bibinfo{year}{2017}).
\newblock \bibinfo{title}{Segnet: A deep convolutional encoder-decoder
  architecture for image segmentation}.
\newblock {\it \bibinfo{journal}{IEEE transactions on pattern analysis and
  machine intelligence}\/},  {\it \bibinfo{volume}{39}\/},
  \bibinfo{pages}{2481--2495}.
\bibitem[{Barata et~al.(2021)Barata, Celebi \& Marques}]{barata2021explainable}
\bibinfo{author}{Barata, C.}, \bibinfo{author}{Celebi, M.~E.}, \&
  \bibinfo{author}{Marques, J.~S.} (\bibinfo{year}{2021}).
\newblock \bibinfo{title}{Explainable skin lesion diagnosis using taxonomies}.
\newblock {\it \bibinfo{journal}{Pattern Recognition}\/},  {\it
  \bibinfo{volume}{110}\/}, \bibinfo{pages}{107413}.
\bibitem[{Basak \& Kundu(2020)}]{basak2020comparative}
\bibinfo{author}{Basak, H.}, \& \bibinfo{author}{Kundu, R.}
  (\bibinfo{year}{2020}).
\newblock \bibinfo{title}{Comparative study of maturation profiles of neural
  cells in different species with the help of computer vision and deep
  learning}.
\newblock In {\it \bibinfo{booktitle}{International Symposium on Signal
  Processing and Intelligent Recognition Systems}\/} (pp.
  \bibinfo{pages}{352--366}).
\newblock \bibinfo{organization}{Springer}.
\bibitem[{Beuren et~al.(2012)Beuren, Janasieivicz, Pinheiro, Grando \&
  Facon}]{beuren2012skin}
\bibinfo{author}{Beuren, A.~T.}, \bibinfo{author}{Janasieivicz, R.},
  \bibinfo{author}{Pinheiro, G.}, \bibinfo{author}{Grando, N.}, \&
  \bibinfo{author}{Facon, J.} (\bibinfo{year}{2012}).
\newblock \bibinfo{title}{Skin melanoma segmentation by morphological
  approach}.
\newblock In {\it \bibinfo{booktitle}{Proceedings of the international
  conference on advances in computing, communications and informatics}\/} (pp.
  \bibinfo{pages}{972--978}).
\bibitem[{Bi et~al.(2020)Bi, Feng, Fulham \& Kim}]{bi2020multi}
\bibinfo{author}{Bi, L.}, \bibinfo{author}{Feng, D.~D.},
  \bibinfo{author}{Fulham, M.}, \& \bibinfo{author}{Kim, J.}
  (\bibinfo{year}{2020}).
\newblock \bibinfo{title}{Multi-label classification of multi-modality skin
  lesion via hyper-connected convolutional neural network}.
\newblock {\it \bibinfo{journal}{Pattern Recognition}\/},  {\it
  \bibinfo{volume}{107}\/}, \bibinfo{pages}{107502}.
\bibitem[{Bi et~al.(2019)Bi, Kim, Ahn, Kumar, Feng \& Fulham}]{bi2019step}
\bibinfo{author}{Bi, L.}, \bibinfo{author}{Kim, J.}, \bibinfo{author}{Ahn, E.},
  \bibinfo{author}{Kumar, A.}, \bibinfo{author}{Feng, D.}, \&
  \bibinfo{author}{Fulham, M.} (\bibinfo{year}{2019}).
\newblock \bibinfo{title}{Step-wise integration of deep class-specific learning
  for dermoscopic image segmentation}.
\newblock {\it \bibinfo{journal}{Pattern recognition}\/},  {\it
  \bibinfo{volume}{85}\/}, \bibinfo{pages}{78--89}.
\bibitem[{Bi et~al.(2017)Bi, Kim, Ahn, Kumar, Fulham \&
  Feng}]{bi2017dermoscopic}
\bibinfo{author}{Bi, L.}, \bibinfo{author}{Kim, J.}, \bibinfo{author}{Ahn, E.},
  \bibinfo{author}{Kumar, A.}, \bibinfo{author}{Fulham, M.}, \&
  \bibinfo{author}{Feng, D.} (\bibinfo{year}{2017}).
\newblock \bibinfo{title}{Dermoscopic image segmentation via multistage fully
  convolutional networks}.
\newblock {\it \bibinfo{journal}{IEEE Transactions on Biomedical
  Engineering}\/},  {\it \bibinfo{volume}{64}\/}, \bibinfo{pages}{2065--2074}.
\bibitem[{Bissoto et~al.(2018)Bissoto, Perez, Ribeiro, Fornaciali, Avila \&
  Valle}]{bissoto2018deep}
\bibinfo{author}{Bissoto, A.}, \bibinfo{author}{Perez, F.},
  \bibinfo{author}{Ribeiro, V.}, \bibinfo{author}{Fornaciali, M.},
  \bibinfo{author}{Avila, S.}, \& \bibinfo{author}{Valle, E.}
  (\bibinfo{year}{2018}).
\newblock \bibinfo{title}{Deep-learning ensembles for skin-lesion segmentation,
  analysis, classification: Recod titans at isic challenge 2018}.
\newblock {\it \bibinfo{journal}{arXiv preprint arXiv:1808.08480}\/}, .
\bibitem[{Burdick et~al.(2018)Burdick, Marques, Weinthal \&
  Furht}]{burdick2018rethinking}
\bibinfo{author}{Burdick, J.}, \bibinfo{author}{Marques, O.},
  \bibinfo{author}{Weinthal, J.}, \& \bibinfo{author}{Furht, B.}
  (\bibinfo{year}{2018}).
\newblock \bibinfo{title}{Rethinking skin lesion segmentation in a
  convolutional classifier}.
\newblock {\it \bibinfo{journal}{Journal of digital imaging}\/},  {\it
  \bibinfo{volume}{31}\/}, \bibinfo{pages}{435--440}.
\bibitem[{Chatterjee et~al.(2019)Chatterjee, Dey \&
  Munshi}]{chatterjee2019integration}
\bibinfo{author}{Chatterjee, S.}, \bibinfo{author}{Dey, D.}, \&
  \bibinfo{author}{Munshi, S.} (\bibinfo{year}{2019}).
\newblock \bibinfo{title}{Integration of morphological preprocessing and
  fractal based feature extraction with recursive feature elimination for skin
  lesion types classification}.
\newblock {\it \bibinfo{journal}{Computer methods and programs in
  biomedicine}\/},  {\it \bibinfo{volume}{178}\/}, \bibinfo{pages}{201--218}.
\bibitem[{Chattopadhyay \& Basak(2020)}]{chattopadhyay2020multi}
\bibinfo{author}{Chattopadhyay, S.}, \& \bibinfo{author}{Basak, H.}
  (\bibinfo{year}{2020}).
\newblock \bibinfo{title}{Multi-scale attention u-net (msaunet): A modified
  u-net architecture for scene segmentation}.
\newblock {\it \bibinfo{journal}{arXiv preprint arXiv:2009.06911}\/}, .
\bibitem[{Chen et~al.(2018)Chen, Tan, Wang \& Hu}]{chen2018reverse}
\bibinfo{author}{Chen, S.}, \bibinfo{author}{Tan, X.}, \bibinfo{author}{Wang,
  B.}, \& \bibinfo{author}{Hu, X.} (\bibinfo{year}{2018}).
\newblock \bibinfo{title}{Reverse attention for salient object detection}.
\newblock In {\it \bibinfo{booktitle}{Proceedings of the European Conference on
  Computer Vision (ECCV)}\/} (pp. \bibinfo{pages}{234--250}).
\bibitem[{Codella et~al.(2018)Codella, Gutman, Celebi, Helba, Marchetti, Dusza,
  Kalloo, Liopyris, Mishra, Kittler et~al.}]{codella2018skin}
\bibinfo{author}{Codella, N.~C.}, \bibinfo{author}{Gutman, D.},
  \bibinfo{author}{Celebi, M.~E.}, \bibinfo{author}{Helba, B.},
  \bibinfo{author}{Marchetti, M.~A.}, \bibinfo{author}{Dusza, S.~W.},
  \bibinfo{author}{Kalloo, A.}, \bibinfo{author}{Liopyris, K.},
  \bibinfo{author}{Mishra, N.}, \bibinfo{author}{Kittler, H.} et~al.
  (\bibinfo{year}{2018}).
\newblock \bibinfo{title}{Skin lesion analysis toward melanoma detection: A
  challenge at the 2017 international symposium on biomedical imaging (isbi),
  hosted by the international skin imaging collaboration (isic)}.
\newblock In {\it \bibinfo{booktitle}{2018 IEEE 15th International Symposium on
  Biomedical Imaging (ISBI 2018)}\/} (pp. \bibinfo{pages}{168--172}).
\newblock \bibinfo{organization}{IEEE}.
\bibitem[{Fabbri et~al.(2008)Fabbri, Costa, Torelli \& Bruno}]{fabbri20082d}
\bibinfo{author}{Fabbri, R.}, \bibinfo{author}{Costa, L. D.~F.},
  \bibinfo{author}{Torelli, J.~C.}, \& \bibinfo{author}{Bruno, O.~M.}
  (\bibinfo{year}{2008}).
\newblock \bibinfo{title}{2d euclidean distance transform algorithms: A
  comparative survey}.
\newblock {\it \bibinfo{journal}{ACM Computing Surveys (CSUR)}\/},  {\it
  \bibinfo{volume}{40}\/}, \bibinfo{pages}{1--44}.
\bibitem[{Fu et~al.(2019)Fu, Liu, Tian, Li, Bao, Fang \& Lu}]{fu2019dual}
\bibinfo{author}{Fu, J.}, \bibinfo{author}{Liu, J.}, \bibinfo{author}{Tian,
  H.}, \bibinfo{author}{Li, Y.}, \bibinfo{author}{Bao, Y.},
  \bibinfo{author}{Fang, Z.}, \& \bibinfo{author}{Lu, H.}
  (\bibinfo{year}{2019}).
\newblock \bibinfo{title}{Dual attention network for scene segmentation}.
\newblock In {\it \bibinfo{booktitle}{Proceedings of the IEEE/CVF Conference on
  Computer Vision and Pattern Recognition}\/} (pp.
  \bibinfo{pages}{3146--3154}).
\bibitem[{Gao et~al.(2019)Gao, Cheng, Zhao, Zhang, Yang \&
  Torr}]{gao2019res2net}
\bibinfo{author}{Gao, S.}, \bibinfo{author}{Cheng, M.-M.},
  \bibinfo{author}{Zhao, K.}, \bibinfo{author}{Zhang, X.-Y.},
  \bibinfo{author}{Yang, M.-H.}, \& \bibinfo{author}{Torr, P.~H.}
  (\bibinfo{year}{2019}).
\newblock \bibinfo{title}{Res2net: A new multi-scale backbone architecture}.
\newblock {\it \bibinfo{journal}{IEEE transactions on pattern analysis and
  machine intelligence}\/}, .
\bibitem[{Garcia-Arroyo \& Garcia-Zapirain(2019)}]{garcia2019segmentation}
\bibinfo{author}{Garcia-Arroyo, J.~L.}, \& \bibinfo{author}{Garcia-Zapirain,
  B.} (\bibinfo{year}{2019}).
\newblock \bibinfo{title}{Segmentation of skin lesions in dermoscopy images
  using fuzzy classification of pixels and histogram thresholding}.
\newblock {\it \bibinfo{journal}{Computer methods and programs in
  biomedicine}\/},  {\it \bibinfo{volume}{168}\/}, \bibinfo{pages}{11--19}.
\bibitem[{Goyal et~al.(2019)Goyal, Oakley, Bansal, Dancey \&
  Yap}]{goyal2019skin}
\bibinfo{author}{Goyal, M.}, \bibinfo{author}{Oakley, A.},
  \bibinfo{author}{Bansal, P.}, \bibinfo{author}{Dancey, D.}, \&
  \bibinfo{author}{Yap, M.~H.} (\bibinfo{year}{2019}).
\newblock \bibinfo{title}{Skin lesion segmentation in dermoscopic images with
  ensemble deep learning methods}.
\newblock {\it \bibinfo{journal}{IEEE Access}\/},  {\it \bibinfo{volume}{8}\/},
  \bibinfo{pages}{4171--4181}.
\bibitem[{Gu et~al.(2019)Gu, Cheng, Fu, Zhou, Hao, Zhao, Zhang, Gao \&
  Liu}]{gu2019net}
\bibinfo{author}{Gu, Z.}, \bibinfo{author}{Cheng, J.}, \bibinfo{author}{Fu,
  H.}, \bibinfo{author}{Zhou, K.}, \bibinfo{author}{Hao, H.},
  \bibinfo{author}{Zhao, Y.}, \bibinfo{author}{Zhang, T.},
  \bibinfo{author}{Gao, S.}, \& \bibinfo{author}{Liu, J.}
  (\bibinfo{year}{2019}).
\newblock \bibinfo{title}{Ce-net: Context encoder network for 2d medical image
  segmentation}.
\newblock {\it \bibinfo{journal}{IEEE transactions on medical imaging}\/},
  {\it \bibinfo{volume}{38}\/}, \bibinfo{pages}{2281--2292}.
\bibitem[{Hasan et~al.(2020)Hasan, Dahal, Samarakoon, Tushar \&
  Mart{\'\i}}]{hasan2020dsnet}
\bibinfo{author}{Hasan, M.~K.}, \bibinfo{author}{Dahal, L.},
  \bibinfo{author}{Samarakoon, P.~N.}, \bibinfo{author}{Tushar, F.~I.}, \&
  \bibinfo{author}{Mart{\'\i}, R.} (\bibinfo{year}{2020}).
\newblock \bibinfo{title}{Dsnet: Automatic dermoscopic skin lesion
  segmentation}.
\newblock {\it \bibinfo{journal}{Computers in Biology and Medicine}\/},  {\it
  \bibinfo{volume}{120}\/}, \bibinfo{pages}{103738}.
\bibitem[{Ibtehaz \& Rahman(2020)}]{ibtehaz2020multiresunet}
\bibinfo{author}{Ibtehaz, N.}, \& \bibinfo{author}{Rahman, M.~S.}
  (\bibinfo{year}{2020}).
\newblock \bibinfo{title}{Multiresunet: Rethinking the u-net architecture for
  multimodal biomedical image segmentation}.
\newblock {\it \bibinfo{journal}{Neural Networks}\/},  {\it
  \bibinfo{volume}{121}\/}, \bibinfo{pages}{74--87}.
\bibitem[{Jha et~al.(2020)Jha, Riegler, Johansen, Halvorsen \&
  Johansen}]{jha2020doubleu}
\bibinfo{author}{Jha, D.}, \bibinfo{author}{Riegler, M.~A.},
  \bibinfo{author}{Johansen, D.}, \bibinfo{author}{Halvorsen, P.}, \&
  \bibinfo{author}{Johansen, H.~D.} (\bibinfo{year}{2020}).
\newblock \bibinfo{title}{Doubleu-net: A deep convolutional neural network for
  medical image segmentation}.
\newblock In {\it \bibinfo{booktitle}{2020 IEEE 33rd International Symposium on
  Computer-Based Medical Systems (CBMS)}\/} (pp. \bibinfo{pages}{558--564}).
\newblock \bibinfo{organization}{IEEE}.
\bibitem[{Lei et~al.(2020)Lei, Xia, Jiang, Jiang, Ge, Xu, Qin, Chen, Wang \&
  Wang}]{lei2020skin}
\bibinfo{author}{Lei, B.}, \bibinfo{author}{Xia, Z.}, \bibinfo{author}{Jiang,
  F.}, \bibinfo{author}{Jiang, X.}, \bibinfo{author}{Ge, Z.},
  \bibinfo{author}{Xu, Y.}, \bibinfo{author}{Qin, J.}, \bibinfo{author}{Chen,
  S.}, \bibinfo{author}{Wang, T.}, \& \bibinfo{author}{Wang, S.}
  (\bibinfo{year}{2020}).
\newblock \bibinfo{title}{Skin lesion segmentation via generative adversarial
  networks with dual discriminators}.
\newblock {\it \bibinfo{journal}{Medical Image Analysis}\/},  {\it
  \bibinfo{volume}{64}\/}, \bibinfo{pages}{101716}.
\bibitem[{Li et~al.(2021)Li, Ma, Yi, Chen \& Ma}]{li2021single}
\bibinfo{author}{Li, X.}, \bibinfo{author}{Ma, H.}, \bibinfo{author}{Yi, S.},
  \bibinfo{author}{Chen, Y.}, \& \bibinfo{author}{Ma, H.}
  (\bibinfo{year}{2021}).
\newblock \bibinfo{title}{Single annotated pixel based weakly supervised
  semantic segmentation under driving scenes}.
\newblock {\it \bibinfo{journal}{Pattern Recognition}\/},  (p.
  \bibinfo{pages}{107979}).
\bibitem[{Liu et~al.(2018)Liu, Huang et~al.}]{liu2018receptive}
\bibinfo{author}{Liu, S.}, \bibinfo{author}{Huang, D.} et~al.
  (\bibinfo{year}{2018}).
\newblock \bibinfo{title}{Receptive field block net for accurate and fast
  object detection}.
\newblock In {\it \bibinfo{booktitle}{Proceedings of the European Conference on
  Computer Vision (ECCV)}\/} (pp. \bibinfo{pages}{385--400}).
\bibitem[{Long et~al.(2015)Long, Shelhamer \& Darrell}]{long2015fully}
\bibinfo{author}{Long, J.}, \bibinfo{author}{Shelhamer, E.}, \&
  \bibinfo{author}{Darrell, T.} (\bibinfo{year}{2015}).
\newblock \bibinfo{title}{Fully convolutional networks for semantic
  segmentation}.
\newblock In {\it \bibinfo{booktitle}{Proceedings of the IEEE conference on
  computer vision and pattern recognition}\/} (pp.
  \bibinfo{pages}{3431--3440}).
\bibitem[{Ma \& Tavares(2015)}]{ma2015novel}
\bibinfo{author}{Ma, Z.}, \& \bibinfo{author}{Tavares, J. M.~R.}
  (\bibinfo{year}{2015}).
\newblock \bibinfo{title}{A novel approach to segment skin lesions in
  dermoscopic images based on a deformable model}.
\newblock {\it \bibinfo{journal}{IEEE journal of biomedical and health
  informatics}\/},  {\it \bibinfo{volume}{20}\/}, \bibinfo{pages}{615--623}.
\bibitem[{Mahbod et~al.(2020)Mahbod, Tschandl, Langs, Ecker \&
  Ellinger}]{mahbod2020effects}
\bibinfo{author}{Mahbod, A.}, \bibinfo{author}{Tschandl, P.},
  \bibinfo{author}{Langs, G.}, \bibinfo{author}{Ecker, R.}, \&
  \bibinfo{author}{Ellinger, I.} (\bibinfo{year}{2020}).
\newblock \bibinfo{title}{The effects of skin lesion segmentation on the
  performance of dermatoscopic image classification}.
\newblock {\it \bibinfo{journal}{Computer Methods and Programs in
  Biomedicine}\/},  {\it \bibinfo{volume}{197}\/}, \bibinfo{pages}{105725}.
\bibitem[{Mendon{\c{c}}a et~al.(2013)Mendon{\c{c}}a, Ferreira, Marques, Marcal
  \& Rozeira}]{mendoncca2013ph}
\bibinfo{author}{Mendon{\c{c}}a, T.}, \bibinfo{author}{Ferreira, P.~M.},
  \bibinfo{author}{Marques, J.~S.}, \bibinfo{author}{Marcal, A.~R.}, \&
  \bibinfo{author}{Rozeira, J.} (\bibinfo{year}{2013}).
\newblock \bibinfo{title}{Ph 2-a dermoscopic image database for research and
  benchmarking}.
\newblock In {\it \bibinfo{booktitle}{2013 35th annual international conference
  of the IEEE engineering in medicine and biology society (EMBC)}\/} (pp.
  \bibinfo{pages}{5437--5440}).
\newblock \bibinfo{organization}{IEEE}.
\bibitem[{Navarro et~al.(2018)Navarro, Escudero-Vi{\~n}olo \&
  Besc{\'o}s}]{navarro2018accurate}
\bibinfo{author}{Navarro, F.}, \bibinfo{author}{Escudero-Vi{\~n}olo, M.}, \&
  \bibinfo{author}{Besc{\'o}s, J.} (\bibinfo{year}{2018}).
\newblock \bibinfo{title}{Accurate segmentation and registration of skin lesion
  images to evaluate lesion change}.
\newblock {\it \bibinfo{journal}{IEEE journal of biomedical and health
  informatics}\/},  {\it \bibinfo{volume}{23}\/}, \bibinfo{pages}{501--508}.
\bibitem[{{\"O}zt{\"u}rk \& {\"O}zkaya(2020)}]{ozturk2020skin}
\bibinfo{author}{{\"O}zt{\"u}rk, {\c{S}}.}, \& \bibinfo{author}{{\"O}zkaya, U.}
  (\bibinfo{year}{2020}).
\newblock \bibinfo{title}{Skin lesion segmentation with improved convolutional
  neural network}.
\newblock {\it \bibinfo{journal}{Journal of digital imaging}\/},  {\it
  \bibinfo{volume}{33}\/}, \bibinfo{pages}{958--970}.
\bibitem[{Ronneberger et~al.(2015)Ronneberger, Fischer \&
  Brox}]{ronneberger2015u}
\bibinfo{author}{Ronneberger, O.}, \bibinfo{author}{Fischer, P.}, \&
  \bibinfo{author}{Brox, T.} (\bibinfo{year}{2015}).
\newblock \bibinfo{title}{U-net: Convolutional networks for biomedical image
  segmentation}.
\newblock In {\it \bibinfo{booktitle}{International Conference on Medical image
  computing and computer-assisted intervention}\/} (pp.
  \bibinfo{pages}{234--241}).
\newblock \bibinfo{organization}{Springer}.
\bibitem[{Saha et~al.(2020)Saha, Prasad \& Thabit}]{saha2020leveraging}
\bibinfo{author}{Saha, A.}, \bibinfo{author}{Prasad, P.}, \&
  \bibinfo{author}{Thabit, A.} (\bibinfo{year}{2020}).
\newblock \bibinfo{title}{Leveraging adaptive color augmentation in
  convolutional neural networks for deep skin lesion segmentation}.
\newblock In {\it \bibinfo{booktitle}{2020 IEEE 17th International Symposium on
  Biomedical Imaging (ISBI)}\/} (pp. \bibinfo{pages}{2014--2017}).
\newblock \bibinfo{organization}{IEEE}.
\bibitem[{Salido \& Ruiz(2018)}]{salido2018using}
\bibinfo{author}{Salido, J. A.~A.}, \& \bibinfo{author}{Ruiz, C.}
  (\bibinfo{year}{2018}).
\newblock \bibinfo{title}{Using deep learning to detect melanoma in dermoscopy
  images}.
\newblock {\it \bibinfo{journal}{Int. J. Mach. Learn. Comput}\/},  {\it
  \bibinfo{volume}{8}\/}, \bibinfo{pages}{61--68}.
\bibitem[{Shahin et~al.(2019)Shahin, Amer \& Elattar}]{shahin2019deep}
\bibinfo{author}{Shahin, A.~H.}, \bibinfo{author}{Amer, K.}, \&
  \bibinfo{author}{Elattar, M.~A.} (\bibinfo{year}{2019}).
\newblock \bibinfo{title}{Deep convolutional encoder-decoders with aggregated
  multi-resolution skip connections for skin lesion segmentation}.
\newblock In {\it \bibinfo{booktitle}{2019 IEEE 16th International Symposium on
  Biomedical Imaging (ISBI 2019)}\/} (pp. \bibinfo{pages}{451--454}).
\newblock \bibinfo{organization}{IEEE}.
\bibitem[{Siegel et~al.(2019)Siegel, Miller \& Jemal}]{siegel2019cancer}
\bibinfo{author}{Siegel, R.~L.}, \bibinfo{author}{Miller, K.~D.}, \&
  \bibinfo{author}{Jemal, A.} (\bibinfo{year}{2019}).
\newblock \bibinfo{title}{Cancer statistics, 2019}.
\newblock {\it \bibinfo{journal}{CA: a cancer journal for clinicians}\/},  {\it
  \bibinfo{volume}{69}\/}, \bibinfo{pages}{7--34}.
\bibitem[{Telea(2004)}]{telea2004image}
\bibinfo{author}{Telea, A.} (\bibinfo{year}{2004}).
\newblock \bibinfo{title}{An image inpainting technique based on the fast
  marching method}.
\newblock {\it \bibinfo{journal}{Journal of graphics tools}\/},  {\it
  \bibinfo{volume}{9}\/}, \bibinfo{pages}{23--34}.
\bibitem[{Tschandl et~al.(2018)Tschandl, Rosendahl \&
  Kittler}]{tschandl2018ham10000}
\bibinfo{author}{Tschandl, P.}, \bibinfo{author}{Rosendahl, C.}, \&
  \bibinfo{author}{Kittler, H.} (\bibinfo{year}{2018}).
\newblock \bibinfo{title}{The ham10000 dataset, a large collection of
  multi-source dermatoscopic images of common pigmented skin lesions}.
\newblock {\it \bibinfo{journal}{Scientific data}\/},  {\it
  \bibinfo{volume}{5}\/}, \bibinfo{pages}{1--9}.
\bibitem[{Tschandl et~al.(2019)Tschandl, Sinz \& Kittler}]{tschandl2019domain}
\bibinfo{author}{Tschandl, P.}, \bibinfo{author}{Sinz, C.}, \&
  \bibinfo{author}{Kittler, H.} (\bibinfo{year}{2019}).
\newblock \bibinfo{title}{Domain-specific classification-pretrained fully
  convolutional network encoders for skin lesion segmentation}.
\newblock {\it \bibinfo{journal}{Computers in biology and medicine}\/},  {\it
  \bibinfo{volume}{104}\/}, \bibinfo{pages}{111--116}.
\bibitem[{{\"U}nver \& Ayan(2019)}]{unver2019skin}
\bibinfo{author}{{\"U}nver, H.~M.}, \& \bibinfo{author}{Ayan, E.}
  (\bibinfo{year}{2019}).
\newblock \bibinfo{title}{Skin lesion segmentation in dermoscopic images with
  combination of yolo and grabcut algorithm}.
\newblock {\it \bibinfo{journal}{Diagnostics}\/},  {\it \bibinfo{volume}{9}\/},
  \bibinfo{pages}{72}.
\bibitem[{Vasconcelos et~al.(2019)Vasconcelos, Medeiros, Peixoto \&
  Reboucas~Filho}]{vasconcelos2019automatic}
\bibinfo{author}{Vasconcelos, F. F.~X.}, \bibinfo{author}{Medeiros, A.~G.},
  \bibinfo{author}{Peixoto, S.~A.}, \& \bibinfo{author}{Reboucas~Filho, P.~P.}
  (\bibinfo{year}{2019}).
\newblock \bibinfo{title}{Automatic skin lesions segmentation based on a new
  morphological approach via geodesic active contour}.
\newblock {\it \bibinfo{journal}{Cognitive Systems Research}\/},  {\it
  \bibinfo{volume}{55}\/}, \bibinfo{pages}{44--59}.
\bibitem[{Verma et~al.(2015)Verma, Singh \& Thoke}]{verma2015enhancement}
\bibinfo{author}{Verma, K.}, \bibinfo{author}{Singh, B.~K.}, \&
  \bibinfo{author}{Thoke, A.} (\bibinfo{year}{2015}).
\newblock \bibinfo{title}{An enhancement in adaptive median filter for edge
  preservation}.
\newblock {\it \bibinfo{journal}{Procedia Computer Science}\/},  {\it
  \bibinfo{volume}{48}\/}, \bibinfo{pages}{29--36}.
\bibitem[{Wang et~al.(2014)Wang, Wang, Li, Chen, Lu, Ma, Peng, Wang \&
  Tang}]{wang2014morphological}
\bibinfo{author}{Wang, G.}, \bibinfo{author}{Wang, Y.}, \bibinfo{author}{Li,
  H.}, \bibinfo{author}{Chen, X.}, \bibinfo{author}{Lu, H.},
  \bibinfo{author}{Ma, Y.}, \bibinfo{author}{Peng, C.}, \bibinfo{author}{Wang,
  Y.}, \& \bibinfo{author}{Tang, L.} (\bibinfo{year}{2014}).
\newblock \bibinfo{title}{Morphological background detection and illumination
  normalization of text image with poor lighting}.
\newblock {\it \bibinfo{journal}{PLoS One}\/},  {\it \bibinfo{volume}{9}\/},
  \bibinfo{pages}{e110991}.
\bibitem[{Wang et~al.(2019)Wang, Wang, Sheng \& Zhang}]{wang2019automated}
\bibinfo{author}{Wang, H.}, \bibinfo{author}{Wang, G.}, \bibinfo{author}{Sheng,
  Z.}, \& \bibinfo{author}{Zhang, S.} (\bibinfo{year}{2019}).
\newblock \bibinfo{title}{Automated segmentation of skin lesion based on
  pyramid attention network}.
\newblock In {\it \bibinfo{booktitle}{International Workshop on Machine
  Learning in Medical Imaging}\/} (pp. \bibinfo{pages}{435--443}).
\newblock \bibinfo{organization}{Springer}.
\bibitem[{Wei et~al.(2017)Wei, Feng, Liang, Cheng, Zhao \& Yan}]{wei2017object}
\bibinfo{author}{Wei, Y.}, \bibinfo{author}{Feng, J.}, \bibinfo{author}{Liang,
  X.}, \bibinfo{author}{Cheng, M.-M.}, \bibinfo{author}{Zhao, Y.}, \&
  \bibinfo{author}{Yan, S.} (\bibinfo{year}{2017}).
\newblock \bibinfo{title}{Object region mining with adversarial erasing: A
  simple classification to semantic segmentation approach}.
\newblock In {\it \bibinfo{booktitle}{Proceedings of the IEEE conference on
  computer vision and pattern recognition}\/} (pp.
  \bibinfo{pages}{1568--1576}).
\bibitem[{Wei et~al.(2019)Wei, Song, Chen, Li \& Han}]{wei2019attention}
\bibinfo{author}{Wei, Z.}, \bibinfo{author}{Song, H.}, \bibinfo{author}{Chen,
  L.}, \bibinfo{author}{Li, Q.}, \& \bibinfo{author}{Han, G.}
  (\bibinfo{year}{2019}).
\newblock \bibinfo{title}{Attention-based denseunet network with adversarial
  training for skin lesion segmentation}.
\newblock {\it \bibinfo{journal}{IEEE Access}\/},  {\it \bibinfo{volume}{7}\/},
  \bibinfo{pages}{136616--136629}.
\bibitem[{Weng et~al.(2019)Weng, Zhou, Li \& Qiu}]{weng2019unet}
\bibinfo{author}{Weng, Y.}, \bibinfo{author}{Zhou, T.}, \bibinfo{author}{Li,
  Y.}, \& \bibinfo{author}{Qiu, X.} (\bibinfo{year}{2019}).
\newblock \bibinfo{title}{Nas-unet: Neural architecture search for medical
  image segmentation}.
\newblock {\it \bibinfo{journal}{IEEE Access}\/},  {\it \bibinfo{volume}{7}\/},
  \bibinfo{pages}{44247--44257}.
\bibitem[{Wu et~al.(2019)Wu, Su \& Huang}]{wu2019cascaded}
\bibinfo{author}{Wu, Z.}, \bibinfo{author}{Su, L.}, \& \bibinfo{author}{Huang,
  Q.} (\bibinfo{year}{2019}).
\newblock \bibinfo{title}{Cascaded partial decoder for fast and accurate
  salient object detection}.
\newblock In {\it \bibinfo{booktitle}{Proceedings of the IEEE/CVF Conference on
  Computer Vision and Pattern Recognition}\/} (pp.
  \bibinfo{pages}{3907--3916}).
\bibitem[{Xie et~al.(2020)Xie, Yang, Liu, Jiang, Zheng \& Wang}]{xie2020skin}
\bibinfo{author}{Xie, F.}, \bibinfo{author}{Yang, J.}, \bibinfo{author}{Liu,
  J.}, \bibinfo{author}{Jiang, Z.}, \bibinfo{author}{Zheng, Y.}, \&
  \bibinfo{author}{Wang, Y.} (\bibinfo{year}{2020}).
\newblock \bibinfo{title}{Skin lesion segmentation using high-resolution
  convolutional neural network}.
\newblock {\it \bibinfo{journal}{Computer methods and programs in
  biomedicine}\/},  {\it \bibinfo{volume}{186}\/}, \bibinfo{pages}{105241}.
\bibitem[{Yuan et~al.(2017)Yuan, Chao \& Lo}]{yuan2017automatic}
\bibinfo{author}{Yuan, Y.}, \bibinfo{author}{Chao, M.}, \& \bibinfo{author}{Lo,
  Y.-C.} (\bibinfo{year}{2017}).
\newblock \bibinfo{title}{Automatic skin lesion segmentation using deep fully
  convolutional networks with jaccard distance}.
\newblock {\it \bibinfo{journal}{IEEE transactions on medical imaging}\/},
  {\it \bibinfo{volume}{36}\/}, \bibinfo{pages}{1876--1886}.
\bibitem[{Zhang et~al.(2019)Zhang, Fu, Dai, Shen, Pang \& Shao}]{zhang2019net}
\bibinfo{author}{Zhang, Z.}, \bibinfo{author}{Fu, H.}, \bibinfo{author}{Dai,
  H.}, \bibinfo{author}{Shen, J.}, \bibinfo{author}{Pang, Y.}, \&
  \bibinfo{author}{Shao, L.} (\bibinfo{year}{2019}).
\newblock \bibinfo{title}{Et-net: A generic edge-attention guidance network for
  medical image segmentation}.
\newblock In {\it \bibinfo{booktitle}{International Conference on Medical Image
  Computing and Computer-Assisted Intervention}\/} (pp.
  \bibinfo{pages}{442--450}).
\newblock \bibinfo{organization}{Springer}.
\bibitem[{Zhao et~al.(2019)Zhao, Liu, Fan, Cao, Yang \& Cheng}]{zhao2019egnet}
\bibinfo{author}{Zhao, J.-X.}, \bibinfo{author}{Liu, J.-J.},
  \bibinfo{author}{Fan, D.-P.}, \bibinfo{author}{Cao, Y.},
  \bibinfo{author}{Yang, J.}, \& \bibinfo{author}{Cheng, M.-M.}
  (\bibinfo{year}{2019}).
\newblock \bibinfo{title}{Egnet: Edge guidance network for salient object
  detection}.
\newblock In {\it \bibinfo{booktitle}{Proceedings of the IEEE/CVF International
  Conference on Computer Vision}\/} (pp. \bibinfo{pages}{8779--8788}).
\bibitem[{Zhou et~al.(2018)Zhou, Siddiquee, Tajbakhsh \&
  Liang}]{zhou2018unet++}
\bibinfo{author}{Zhou, Z.}, \bibinfo{author}{Siddiquee, M. M.~R.},
  \bibinfo{author}{Tajbakhsh, N.}, \& \bibinfo{author}{Liang, J.}
  (\bibinfo{year}{2018}).
\newblock \bibinfo{title}{Unet++: A nested u-net architecture for medical image
  segmentation}.
\newblock In {\it \bibinfo{booktitle}{Deep learning in medical image analysis
  and multimodal learning for clinical decision support}\/} (pp.
  \bibinfo{pages}{3--11}).
\newblock \bibinfo{publisher}{Springer}.

\end{thebibliography}

\end{document}